\documentclass[prb,aps,twocolumn,showpacs,superscriptaddress]{revtex4-1}
\usepackage{graphics}
\usepackage{epsfig}
\usepackage{amsmath}
\usepackage{amsfonts}
\usepackage{amssymb}
\usepackage{graphicx}
\usepackage{color}
\usepackage{tabularx}
\usepackage{txfonts}

\begin{document}

\title{Pressure dependence of dynamically screened Coulomb interactions in NiO: Effective Hubbard, Hund, intershell and intersite components}
\author{S. K. Panda}
\email{Electronic address: swarup.panda@polytechnique.edu}
\affiliation{Centre de Physique Theorique, Ecole Polytechnique, CNRS UMR 7644, Universite Paris-Saclay, 91128 Palaiseau, France}
\author{H. Jiang}
\affiliation{College of Chemistry and Molecular Engineering, Peking University, 100871 Beijing, China}
\author{S. Biermann}
\email{Electronic address: silke.biermann@polytechnique.edu}
\affiliation{Centre de Physique Theorique, Ecole Polytechnique, CNRS UMR 7644, Universite Paris-Saclay, 91128 Palaiseau, France}
\affiliation{College de France, 11 place Marcelin Berthelot, 75005 Paris, France}
\begin{abstract}
In this work, we report the pressure dependence of the effective Coulomb interaction parameters (Hubbard $U$) in 
paramagnetic
NiO within the constrained random phase approximation (cRPA).  We consider five different low energy models starting from the most expensive one that treats  both Ni-$d$ and O-$p$ states as correlated orbitals ($dp$-$dp$ model) to the smallest possible two-orbital model comprising the $e_g$ states only ($e_g$-$e_g$ model). We find that in all the considered models, the bare interactions are not very sensitive to the compression. However the partially screened interaction parameters show an almost linear increment as a function of compression, resulting from the substantial weakening of screening effects upon compression. This counterintuitive trend is explained from the specific characteristic changes of the basic electronic structure of this system. We further calculate the nearest neighbor inter-site $d$-$d$ interaction terms which also show substantial enhancement due to compression. Our results for both the experimental and highly compressed structures reveal that the frequency dependence of the partially screened interactions can not be ignored in a realistic modeling of NiO. We also find that the computed interaction parameters for the antiferromagnetic NiO are almost identical to their paramagnetic counter parts. 
\end{abstract}

\maketitle

\section{Introduction}
The study of the role of electronic correlations in transition metal compounds is one of the most actively developing fields of modern condensed matter physics.
This class of materials exhibits rich and highly nontrivial physical phenomena such as 
unconventional transport properties including metal-insulator transitions,
bad metal behavior or superconductivity, or ordering phenomena involving
charge, orbital or spin degrees of freedom.
It is by now clear that most of these properties arise due to electronic correlations, in particular strong Coulomb interactions among the partially occupied 3$d$ electrons of the metal ion. 
\par 
Among these oxides, NiO is a prototype system for strong electronic correlations with a high spin antiferromagnetic structure at low temperatures~\cite{NiO_AFM_ordering}. 
The exact origin, nature and size of the fundamental gap and, more broadly, the electronic structure of NiO, has been studied intensely for many years, and is also a major topic of text books on condensed matter physics~\cite{Mott_Orig,Brandow1977,PhysRevLett.52.1830,PhysRevB.30.4734}. Initially NiO was thought to be a classic Mott insulating system~\cite{Mott_Orig,Brandow1977} with a large insulating gap. However, later studies~\cite{PhysRevLett.52.1830,PhysRevB.30.4734} suggested that the gap might open between states of oxygen $p$ character and empty $d$ states, classifying NiO as a charge-transfer insulator.  Over the years a vast number of experimental ~\cite{PhysRevLett.34.395,PhysRevB.26.4845,PhysRevLett.53.2339,PhysRevB.33.4253,PhysRevB.44.3604,PhysRevB.53.10372,PhysRevLett.99.156404,
PhysRevLett.100.206401,Weinen2015} and theoretical~\cite{Brandow1977,PhysRevB.30.957,PhysRevB.33.8896,Sarma1990,PhysRevB.38.3449,PhysRevLett.64.2442,PhysRevB.44.3604,
PhysRevB.48.16929,PhysRevB.50.8257,PhysRevB.62.16392,PhysRevB.74.195114,Takao2008,PhysRevLett.99.156404,PhysRevB.79.235114,
PhysRevLett.109.186401,PhysRevB.91.115105} studies have been carried out aiming at a clearer understanding of the electronic structure and magnetism of this system. 
Recently, there has been a growing interest also in understanding the properties
of NiO under pressure~\cite{PhysRevB.61.14984,Gavrilyuk2001,Cohen654,PhysRevB.69.035114,PhysRevLett.109.086402}. Indeed, a pressure-driven Mott insulator-to-metal transition (IMT) was anticipated by Mott {\em et al.}~\cite{Mott_IMT_NiO} a long time ago, but the compression ration and the corresponding transition pressure $P_{\rm IMT}$ were subjects of debate~\cite{PhysRevB.69.035114}. The Density functional theory calculations using Perdew-Wang generalized gradient approximation (PWGGA) functional~\cite{PhysRevB.69.035114} predict the transition to occur at about 40$\%$ volume compression [(V0-V)/V0], while a calculation using hybrid functional B3LYP~\cite{PhysRevB.69.035114} estimates it to be 65$\%$. These volumes compression correspond to the critical pressure of transition P$_c$ $\sim$ 320 Gpa and $\sim$ 1320 GPa respectively according to their calculated equations of state~\cite{PhysRevB.69.035114} which has been found to be very similar for both the functionals and is also in good agreement with the experimental report of Ref.~\onlinecite{PhysRevB.61.14984}.
The first experimental observation of the transition~\cite{PhysRevLett.109.086402} finally came in 2012, at a transition pressure of 240 GPa. This transition pressure was also calculated based on the above mentioned theoretically calculated equation of state~\cite{PhysRevB.69.035114} and it corresponds to 35$\%$ volume compression. 

On the theoretical side~\cite{PhysRevB.5.290,Cai2009}, it is now well established that neither the conventional effective single-particle band theory within local (spin) density approximation (L(S)DA), nor the L(S)DA+U method in which L(S)DA is augmented by an on-site Coulomb repulsion term and an exchange term with the Hubbard $U$ and Hund exchange $J$ parameters, respectively can provide an accurate description of the electronic properties of NiO. 
Hedin's GW approximation
~\cite{PhysRev.139.A796} 
provides an interesting route to the antiferromagnetic insulating
phase~\cite{PhysRevLett.93.126406,PhysRevB.71.193102,PhysRevB.82.045108} but cannot
describe the paramagnetic insulator~\footnote{This limitation of GW based approach is also true for other transiton metal compounds as shown in Ref.~\onlinecite{PhysRevLett.96.226402,PhysRevLett.102.126403}.}. A more sophisticated treatment of the correlation effects within a fully many-body technique like LDA plus dynamical mean-field theory (LDA+DMFT)~\cite{PhysRevB.57.6884,RevModPhys.68.13} is inevitable to describe the nonquasiparticle features of the electronic spectrum of NiO. 
Even then, NiO turns out to be a tremendous challenge, due to its 
charge transfer character~\cite{PhysRevB.74.195114,PhysRevLett.99.156404,PhysRevB.75.165115,Karolak201011,PhysRevLett.109.186401,PhysRevB.93.235138,Leonov,
NiO_DMFT1,Nekrasov2012,Kunes2009,PhysRevB.75.165115,PhysRevB.77.195124}.
In the LDA+DMFT method, the noninteracting part of the Hamiltonian, obtained from LDA involving a large number of valence $s$, $p$, and $d$ orbitals associated with all the atoms in the unit cell is expressed in an effective $d$-like Wannier basis in order to construct an effective multiband Hubbard model which is solved by DMFT. The interaction term for this Hamiltonian is the effective interaction between the Wannier orbitals which is again incorporated by a Hubbard $U$ and Hund exchange $J$ parameter.
\par
The LDA+DMFT simulations of NiO reported so far~\cite{PhysRevB.74.195114,PhysRevLett.99.156404,PhysRevB.75.165115,Karolak201011,PhysRevLett.109.186401,PhysRevB.93.235138,Leonov,
NiO_DMFT1,Nekrasov2012,Kunes2009,PhysRevB.75.165115,PhysRevB.77.195124} considered $U$ as an adjustable parameter to correctly reproduce the experimentally reported x-ray photoemission spectroscopy (XPS)  or/and bremsstrahlung isochromat spectroscopy (BIS) spectra~\cite{PhysRevLett.34.395,PhysRevLett.53.2339,PhysRevB.33.4253,PhysRevLett.100.206401,Weinen2015}. The different values of $U$ and $J$, used in these studies are tabulated in Table~\ref{UJ_reported}. Values of $U$ ranging from 7.0 eV to 10 eV have been considered.  However there exists a first-principles method, the constrained random-phase approximation (cRPA)~\cite{cRPA_orig,cRPA_wien2kImple} which has been recently successfully employed to evaluate the  screened Coulomb interaction matrix elements ($U$ an $J$) between the effective Wannier functions in many transition metal compounds~\cite{cRPA_wien2kImple, PhysRevB.77.085122,PhysRevB.80.155134,PhysRevB.83.121101,PhysRevLett.107.266404,PhysRevB.86.085117,PhysRevB.86.165124,pandaNiS,PhysRevB.89.125110,PhysRevB.94.125147}, including a study of MnO under pressure \cite{PhysRevB.79.235133,PhysRevB.81.115116}. This method~\cite{cRPA_orig,cRPA_wien2kImple} is capable of efficiently determining all the possible Coulomb matrix elements, e.g., on site, off site, intraorbital, interorbital, and exchange, as well as their frequency dependence~\cite{0953-8984-26-17-173202,PhysRevB.94.125147,Seth-et-al}.  
\par 
Despite enormous work on NiO, the detailed investigation of the Coulomb interaction parameters within the cRPA approach is limited so far. Sakuma {\em et al.}~\cite{PhysRevB.87.165118} reported matrix elements of the partially screened interactions including their frequency dependence using various low energy models and discussed the sensitivity of these parameters with respect to the chosen low energy model~\cite{PhysRevB.87.165118}. 
Seth {\em et al.}~\cite{Seth-et-al} used NiO to benchmark the performance
of a variant of the cRPA scheme dubbed ``shell-folding'', which incorporates
the screening by $pd$-interactions into effective $d$-$d$ and $p$-$p$ interactions.
However, none of these works addressed the important question of how
the effective interactions evolve under pressure.
The effects of external pressure on these very important parameters ($U$ and $J$) of NiO, a most enigmatic strongly correlated material that has been of interest since the days of Mott and Hubbard, has not been analyzed so far. 
A very recent theoretical study~\cite{Leonov} based on the state-of-the-art fully charge self-consistent DFT+DMFT method found the Mott IMT transition to coincide with a magnetic-to-nonmagnetic collapse transition at about 45$\%$ volume compression which corresponds to a critical pressure of  429 GPa according to their calculated equation of state. This study~\cite{Leonov} considered the value of $U$ and $J$ as high as 10 eV and 1 eV respectively and assumed those parameters to remain constant upon variation of the unit cell volume. However the effective $U$ in transition metal compounds may be strongly dependent on pressure as has been emphasized in several previous works~\cite{KunesMnO,Ovchinnikov2008,PhysRevB.79.085125,PhysRevB.81.115116}. The practice of considering $U$ as a nearly pressure independent parameter may thus bias the results altogether. Therefore a quantitative estimation of the interactions parameters under pressure is particularly important for a material as important as NiO whose behavior is primarily dominated by the strong Coulomb interactions. 
\bgroup
\def\arraystretch{1.3}
\begin{table}
\caption{The value of $U$ and $J$ (in eV) used in different reported LDA+DMFT studies.}
\vspace{0.1cm}
\begin{tabular}{| c c c c c c |}
\hline
  & Ref.~\onlinecite{PhysRevB.77.195124} & Ref.~\onlinecite{PhysRevLett.109.186401} &  Ref.~\onlinecite{PhysRevB.74.195114,PhysRevLett.99.156404,PhysRevB.75.165115,Karolak201011,Nekrasov2012,Kunes2009} &  Ref.~\onlinecite{PhysRevB.93.235138} & Ref.~\onlinecite{Leonov} \\[1 ex]
\hline
$U$ & 7.0 & 7.5 & 8.0  & 8.5 & 10.0  \\[1 ex]
$J$ & 0.9 & 1.2 & 1.0 & 0.8 & 1.0   \\[1 ex]
\hline
\end{tabular}
\label{UJ_reported}
\end{table}
\egroup

\bgroup
\def\arraystretch{1.6}
\begin{table}
\caption{The amount of volume compression ([V0-V]/V0) and the corresponding calculated pressure ($P$) as reported in Ref~\onlinecite{PhysRevB.69.035114}.}
\vspace{0.1cm}
\begin{tabular}{| c | c c c c c |}
\hline
\hspace{0.03cm} [V0-V]/V0 ($\%$) \hspace{0.03cm} & \hspace{0.03cm} 0.0 \hspace{0.03cm} & \hspace{0.03cm} 11.7 \hspace{0.03cm} & \hspace{0.03cm} 24.3 \hspace{0.03cm} & \hspace{0.03cm} 35.7\hspace{0.03cm} & \hspace{0.03cm} 40.9\hspace{0.03cm} \\[1 ex]
\hline
\hspace{0.03cm} $P$ (Gpa)\hspace{0.03cm} & \hspace{0.03cm} 0\hspace{0.03cm} & \hspace{0.03cm} 32 \hspace{0.03cm} & \hspace{0.03cm} 106 \hspace{0.03cm} & \hspace{0.03cm} 238 \hspace{0.03cm}  & \hspace{0.03cm} 346 \hspace{0.03cm} \\[1 ex]
\hline
\end{tabular}
\label{vol_pressure}
\end{table}
\egroup
\par 
In light of the above, we employed the cRPA method~\cite{cRPA_orig} within the full-potential linearized augmented-plane-wave (LAPW+lo) framework~\cite{lapwPluslo} to provide a complete description of the  correlation matrix elements both at the static limit and the frequency dependence, under pressure within five different low energy models which are relevant for the DMFT simulations.  Our detailed study reveals that the pressure induced changes in the bare Coulomb interactions are negligible, implying non-significant changes of the spread of the Wannier functions upon compression. However we find that due to the certain modifications of the electronic structure, the different screening channels become stronger which essentially enhance the screened $U$ and $J$ substantially. This establishes that the pressure induced change in the correlation matrix elements can not be ignored in order to provide a reliable physical description of the correlated oxides. 
\par 
The paper is organized as follows. In Sec. II, we briefly discuss  the methodology. In Sec. III, we discuss the evolution of the basic electronic structure due to the compression of the unit cell volume and also report estimated values of the static and dynamically screened Coulomb interaction parameters as well as matrix elements for different volumes of the unit cell. Finally, we summarize our conclusions in Sec. IV.
\begin{figure}[t]
\includegraphics[width=0.99\columnwidth]{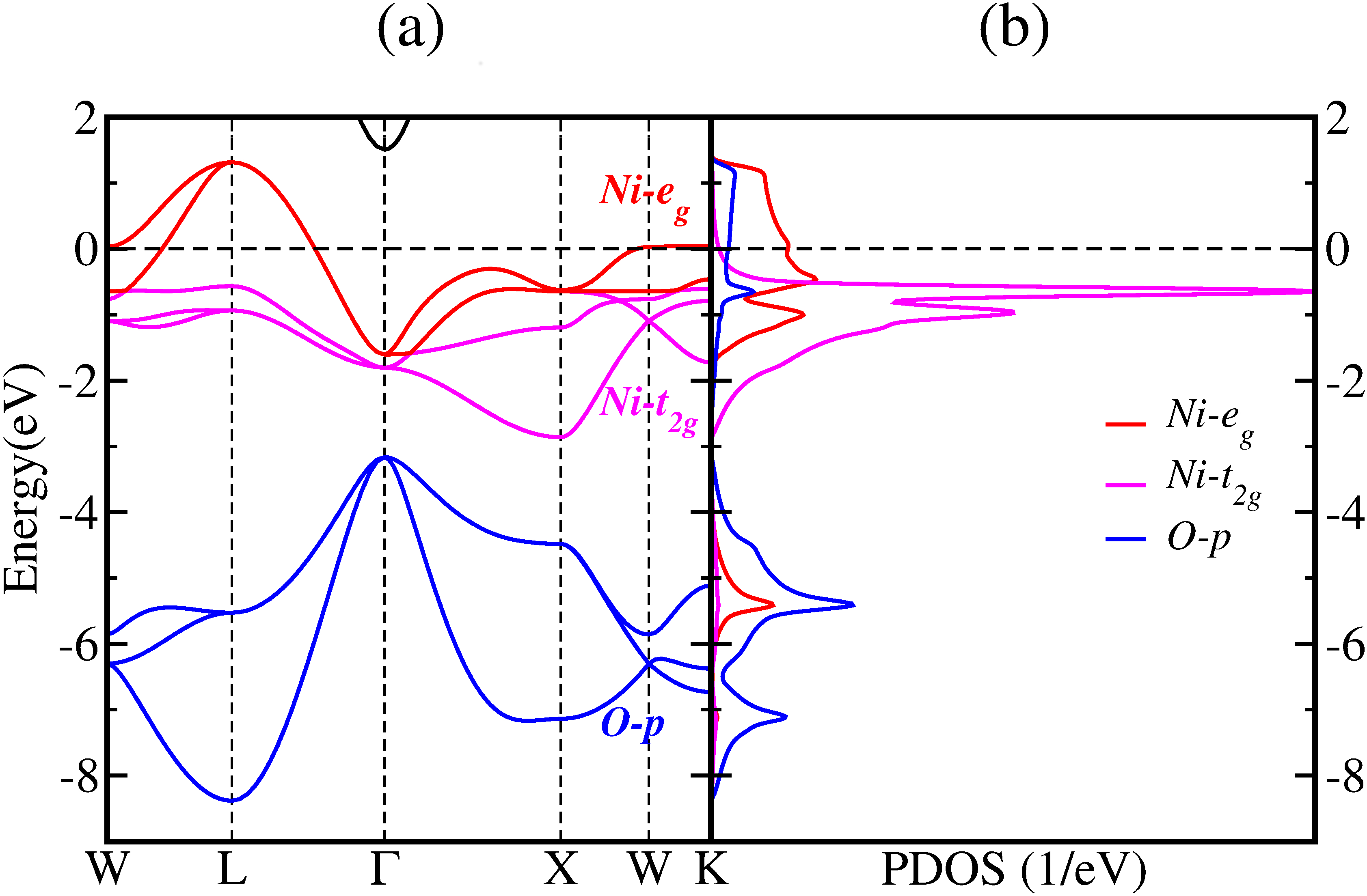}
\caption{(a) The band dispersion and (b) the partial density of states (PDOS) of fcc NiO. Calculations are carried out with experimental lattice parameters.}
\label{bandDOS}
\end{figure}

\begin{figure}[t]
\includegraphics[width=0.99\columnwidth]{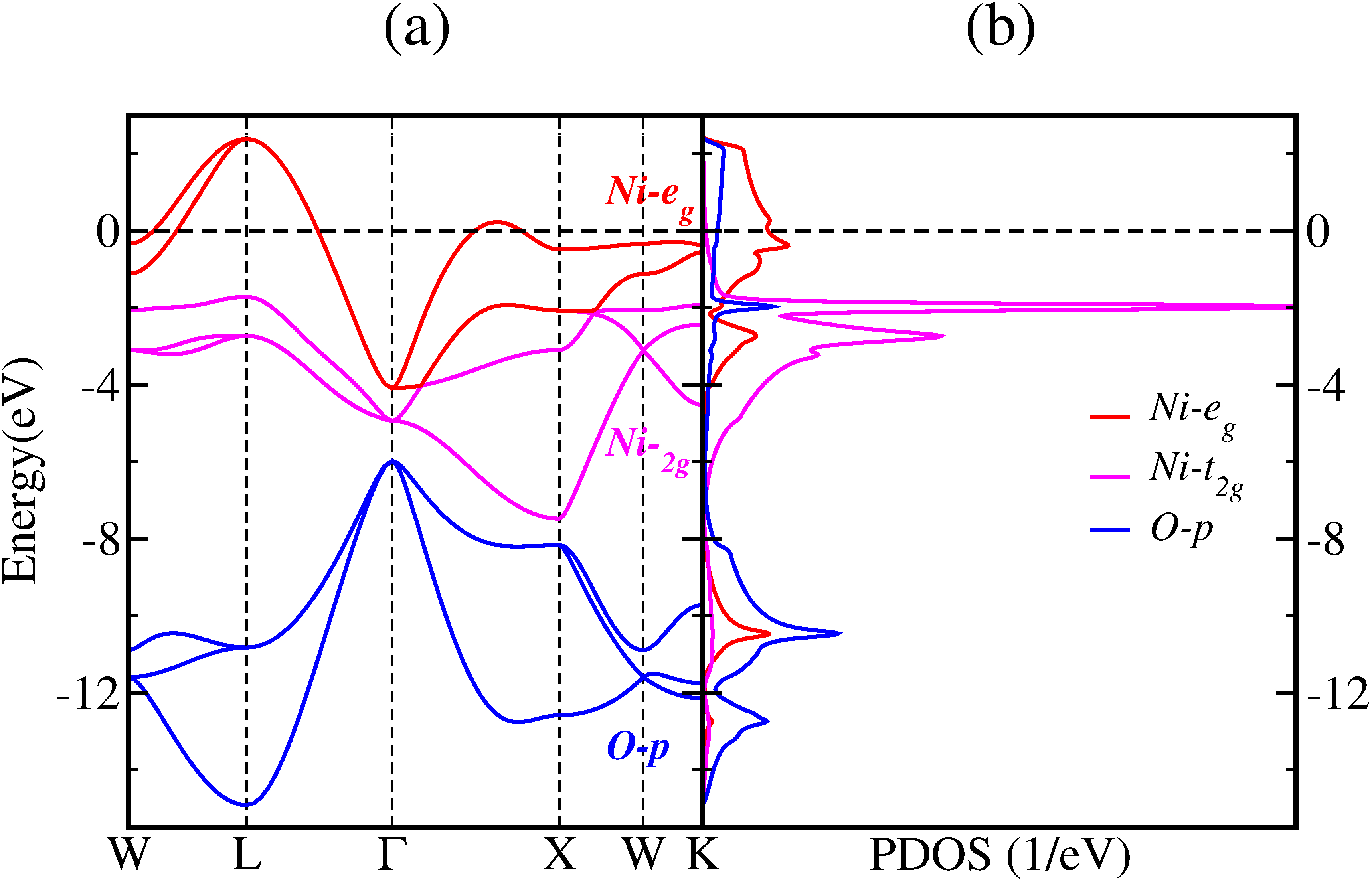}
\caption{(a) The band dispersion and (b) the partial density of states (PDOS) of fcc NiO. Calculations are carried out with the reduced lattice parameters, corresponding to the volume V = 59\%V$_0$, V$_0$ being the experimental volume.}
\label{bandDOS1}
\end{figure}

 \begin{figure*}[t]
\includegraphics[width=1.89\columnwidth]{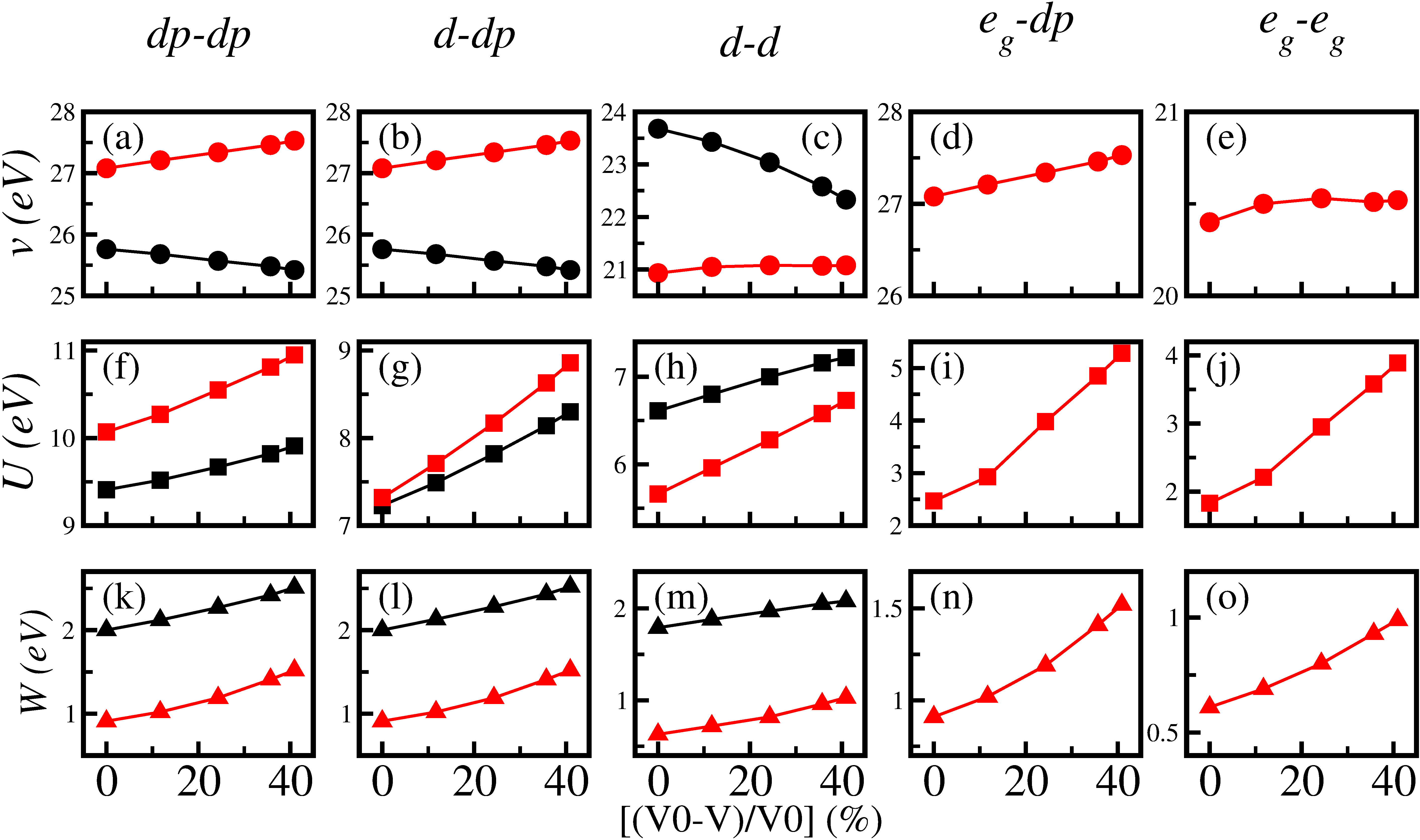}
\caption{Local Coulomb interactions (diagonal element $U_{mm}$ {\it please check!}) of the $t_{2g}$ (black line) and $e_{g}$ (red line) states as a function of the percentage of unit cell volume contraction for the five relevant low energy models. Orbital resolved diagonal elements of the bare interactions $v$ for (a) $dp$-$dp$, (b) $d$-$dp$, (c) $d$-$d$, (d) $e_g$-$dp$, (e) $e_g$-$e_g$ models. Partially screened interaction $U$ for (f) $dp$-$dp$, (g) $d$-$dp$, (h) $d$-$d$, (i) $e_g$-$dp$, (j) $e_g$-$e_g$ models and fully screened interaction $W$ for (k) $dp$-$dp$, (l) $d$-$dp$, (m) $d$-$d$, (n) $e_g$-$dp$, (o) $e_g$-$e_g$ models at zero frequency limit.}
\label{UvW_static}
\end{figure*}

\section{Methodology}
The effective Coulomb interaction parameters and matrix elements, presented in this work are calculated within the recently developed cRPA method~\cite{cRPA_orig} based on the implemention of Ref.~\onlinecite{cRPA_wien2kImple} within the popular Wien2k code~\cite{wien2k}. While an extended description of the technical aspects can be found in Ref.~\onlinecite{cRPA_wien2kImple}, we will briefly describe here the most relevant points and notations for the understanding of our results. 
\par 
A fully charge self-consistent converged DFT-LDA electronic structure is the starting point for the cRPA calculations. 
For this, we carried out density functional theory (DFT)~\cite{DFT1,DFT2} based calculations within local density approximation (LDA) in a non-spin polarized set-up using the all-electron, full potential code WIEN2K~\cite{wien2k}. A dense 30$\times$30$\times$30 k-mesh is considered for the Brillouin-Zone integration. The muffin-tin radii ($R_{MT}$) of Ni and O ions are chosen to be 0.94~\r{A}, and 0.80~\r{A}, respectively for all the lattice parameters. To achieve energy convergence of the eigenvalues, the wave functions in the interstitial region were expanded in plane waves with a cutoff $R_{MT}k_{max}$=8, where $R_{MT}$ denotes the smallest atomic sphere radius and $k_{max}$ represents the magnitude of the largest k vector in the plane wave expansion. The valence wave functions inside the spheres are expanded up to $l_{max}$=10, while the charge density is Fourier expanded up to a large value of $G_{max}$ = 20. In order to simulate the pressure, the calculations have been carried out with the experimental lattice parameters and four other reduced lattice parameters. The percentage of volume compression and the corresponding calculated pressure as reported in Ref~\onlinecite{PhysRevB.69.035114} are shown in Table~\ref{vol_pressure}.
\par 
The first step of our scheme is to construct a set of localized Wannier(-like) orbitals $\{\phi_m\}$ corresponding to the correlated orbitals used in a DFT+DMFT study. The effective partially screened four index Coulomb interactions matrix elements between those Wannier like orbitals can be expressed as
\begin{equation}
U_{m_1,m_2,m_3,m_4} (\omega) = \langle \phi_{m_1}\phi_{m_2} |\varepsilon_r^{-1}(\omega)v| \phi_{m_3}\phi_{m_4}\rangle \: ,
\label{eq:hubbardU} 
\end{equation}
where $v$ is the bare Coulomb interaction and $m_1$, $m_2$, $m_3$, $m_4$ are the orbital quantum numbers. Within the RPA scheme, the dielectric function ($\varepsilon$) is related to the electron irreducible polarizability $P$ by the relation $\varepsilon = 1 - vP$ and the full polarization function ($P$) can be expressed in terms of the Kohn-Sham orbitals ($\psi_{kn}$) and the corresponding eigen values ($\epsilon_{kn}$) as follows.
\begin{align}\nonumber
P (r,r^\prime, \omega) & = \sum_{kn}^{occ}\sum_{k^\prime n^\prime}^{unocc} \psi_{kn}^{\dagger}(r)\psi_{k^\prime n^\prime} (r)\psi_{k^\prime n^\prime}^{\dagger} (r^\prime)\psi_{kn}(r^\prime) \\
& \times \{\frac{1}{\omega - \epsilon_{k^\prime n^\prime} + \epsilon_{kn} + i0^{+}}-\frac{1}{\omega + \epsilon_{k^\prime n^\prime} - \epsilon_{kn} - i0^{+}}\}  \: ,
\label{eq:polarization}
\end{align}
where $n$, $n^\prime$ are the band indices.  Within this formalism, it is also possible to calculate the constrained polarization $P^r = P - P^{sub}$, where we subtract the screening process internal to the correlated subspace ($P^{sub}$) and accordingly we calculate $\varepsilon_r$ which enters into Eqn.~\ref{eq:hubbardU}. The frequency dependence of $U$ arises from $P_r(r,r^\prime, \omega)$. We used a 6$\times$6$\times$6 k-mesh and 7 Rydberg energy cut-off for the unoccupied states in our cRPA calculations. 
\par 
Depending on the choice of an energy window to construct the Wannier function and the screening channel, one can define several models as described in Ref.~\onlinecite{cRPA_wien2kImple}. Although the same convention has been followed in this work, we briefly outline those models for the sake of completeness. In our nomenclature, the first index defines the correlated subspace and thus also indicates the predominant character of the Kohn-Sham bands relevant for calculating $P^{sub}$, while second index defines the band character within the chosen energy window for the construction of the Wannier function. 
\begin{itemize}
 \item $dp$-$dp$ model: In this model both the $d$ and $p$ states are considered to be correlated and all those transitions are removed. Naturally the chosen energy window also consists of states from both the $d$ and $p$ manifold. This model provides the Hubbard $U$ for both $d$ and $p$ states and the interactions between them.
  \item $d$-$dp$ model: This is a hybrid model. Here the Wannier functions are constructed within the extended $dp$-energy window, however it is considered that the Hubbard $U$ is present only for Ni-$d$ states and only those internal transitions are removed from the polarization.  
   \item $d$-$d$ model:  The model is built from the Ni-$d$ like bands where the internal transition within the $d$ bands are removed from the total polarization and the Wannier function is also generated from those bands only. 
\item $e_g$-$dp$ model: This is also another hybrid model, which is similar to the $d$-$dp$ model. Wannier functions are constructed from the $dp$ energy window. However instead of full Ni-$d$ manifold, only the top two $e_g$  states are considered to be correlated and and the corresponding internal transitions are removed from the polarization.  
  \item $e_g$-$e_g$ model: This model is similar to the previous $d$-$d$ model. Here instead of $d$, the model is built from the bands, having predominant Ni-$e_g$ character.
 \end{itemize}
\bgroup
\def\arraystretch{1.2}
\begin{table*}[t]
\caption{On-site $U$ of $t_{2g}$ and $e_{g}$ states and orbital averaged Hund's coupling ($J_{ave}$) of NiO for the two extreme cases (experimental volume V$_0$ and the most compressed volume V = 59\% V$_O$)  corresponding to all the five low energy models. For the $dp$-$dp$ model, we also displayed on-site $U$ of the $p$ states ($U_p$), intersite interactions between the Ni-$d$ and O-$p$ orbitals ($U_{dp}$), and the renormalized $t_{2g}$-$t_{2g}$, $e_{g}$-$e_{g}$ interactions ($U_{t_{2g}}^{SF}$, $U_{e_{g}}^{SF}$) due to the intershell components ($U_{dp}$) following the shell-folding (SF) scheme as introduced in Ref.~\onlinecite{Seth-et-al}.}
\vspace{0.1cm}
\begin{tabular}{| c | c c c c c c c | c c c | c c c | c c | c c |}
\hline
 & \multicolumn{7}{c|}{$dp$-$dp$} & \multicolumn{3}{c|}{$d$-$dp$} & \multicolumn{3}{c|}{$d$-$d$} & \multicolumn{2}{c|}{$e_g$-$dp$} & \multicolumn{2}{c|}{$e_g$-$e_g$}\\
  & \hspace{0.03cm} $U_{t_{2g}}$ \hspace{0.03cm} & \hspace{0.03cm} $U_{e_{g}}$ \hspace{0.03cm} & \hspace{0.03cm}  $U_{p}$ \hspace{0.03cm} & \hspace{0.03cm} $U_{dp}$ \hspace{0.03cm} & \hspace{0.03cm}  $J_{ave}$ \hspace{0.03cm} & \hspace{0.03cm} $U_{t_{2g}}^{SF}$ \hspace{0.03cm} & \hspace{0.03cm} $U_{e_{g}}^{SF}$ \hspace{0.03cm} & \hspace{0.03cm} $U_{t_{2g}}$ \hspace{0.03cm} & \hspace{0.03cm}  $U_{e_{g}}$  \hspace{0.03cm} & \hspace{0.03cm}  $J_{ave}$  \hspace{0.03cm} & \hspace{0.03cm} $U_{t_{2g}}$ \hspace{0.03cm} & \hspace{0.03cm}  $U_{e_{g}}$  \hspace{0.03cm} & \hspace{0.03cm}  $J_{ave}$  \hspace{0.03cm} & \hspace{0.03cm}  $U_{e_{g}}$  \hspace{0.03cm} & \hspace{0.03cm}  $J_{ave}$  \hspace{0.03cm} & \hspace{0.03cm}  $U_{e_{g}}$  \hspace{0.03cm} & \hspace{0.03cm}  $J_{ave}$  \hspace{0.03cm} \\[1 ex]
\hline
V = V$_0$ &  9.41 & 10.07 & 6.78 & 1.18 & 0.85 & 8.23 & 8.89 & 7.23 & 7.32 & 0.84 & 6.61 & 5.66 & 0.69 & 2.47 & 0.72 & 1.83 & 0.52 \\[1 ex]
V = 59\% V$_O$ & 9.91 & 10.95 & 8.31 & 1.55 & 0.86 & 8.36 & 9.40 & 8.30 & 8.86 & 0.85 & 7.22 & 6.73 & 0.66 & 5.28 & 0.96 & 3.89 & 0.66  \\[1 ex]
\hline
\end{tabular}
\label{UJvalues}
\end{table*}
\egroup

\bgroup
\def\arraystretch{1.2}
\begin{table*}
\caption{Averaged nearest-neighbor interatomic bare ($v^{nn}$) and partially screened ($U^{nn}$) interactions between the Ni-$d$ orbitals for the two extreme cases (experimental volume V$_0$ and the most compressed volume V = 59\% V$_O$)  corresponding to all the five low energy models.}
\vspace{0.1cm}
\begin{tabular}{| c | c c | c c | c c | c c | c c |}
\hline
 & \multicolumn{2}{c|}{$dp$-$dp$} & \multicolumn{2}{c|}{$d$-$dp$} & \multicolumn{2}{c|}{$d$-$d$} & \multicolumn{2}{c|}{$e_g$-$dp$} & \multicolumn{2}{c|}{$e_g$-$e_g$}\\
  & \hspace{0.03cm} $v^{nn}$ \hspace{0.03cm} & \hspace{0.03cm} $U^{nn}$ \hspace{0.03cm}  & \hspace{0.03cm} $v^{nn}$ \hspace{0.03cm} & \hspace{0.03cm}  $U^{nn}$ \hspace{0.03cm}  & \hspace{0.03cm} $v^{nn}$ \hspace{0.03cm} & \hspace{0.03cm} $U^{nn}$  \hspace{0.03cm}  & \hspace{0.03cm}  $v^{nn}$  \hspace{0.03cm} & \hspace{0.03cm}  $U^{nn}$  \hspace{0.03cm} & \hspace{0.03cm}  $v^{nn}$  \hspace{0.03cm} & \hspace{0.03cm}  $U^{nn}$  \hspace{0.03cm} \\[1 ex]
\hline
V = V$_0$ &  4.92 & 1.53  & 4.92 & 1.22 & 4.83 & 1.20 & 4.90 & 0.27 & 4.76 & 0.28 \\[1 ex]
V = 59\% V$_O$ & 5.86 & 1.94 & 5.86 & 1.67 & 5.72 & 1.63 & 5.83 & 0.78 & 5.66 & 0.79  \\[1 ex]
\hline
\end{tabular}
\label{Uv_Inter}
\end{table*}
\egroup
\begin{figure}
\includegraphics[width=0.99\columnwidth]{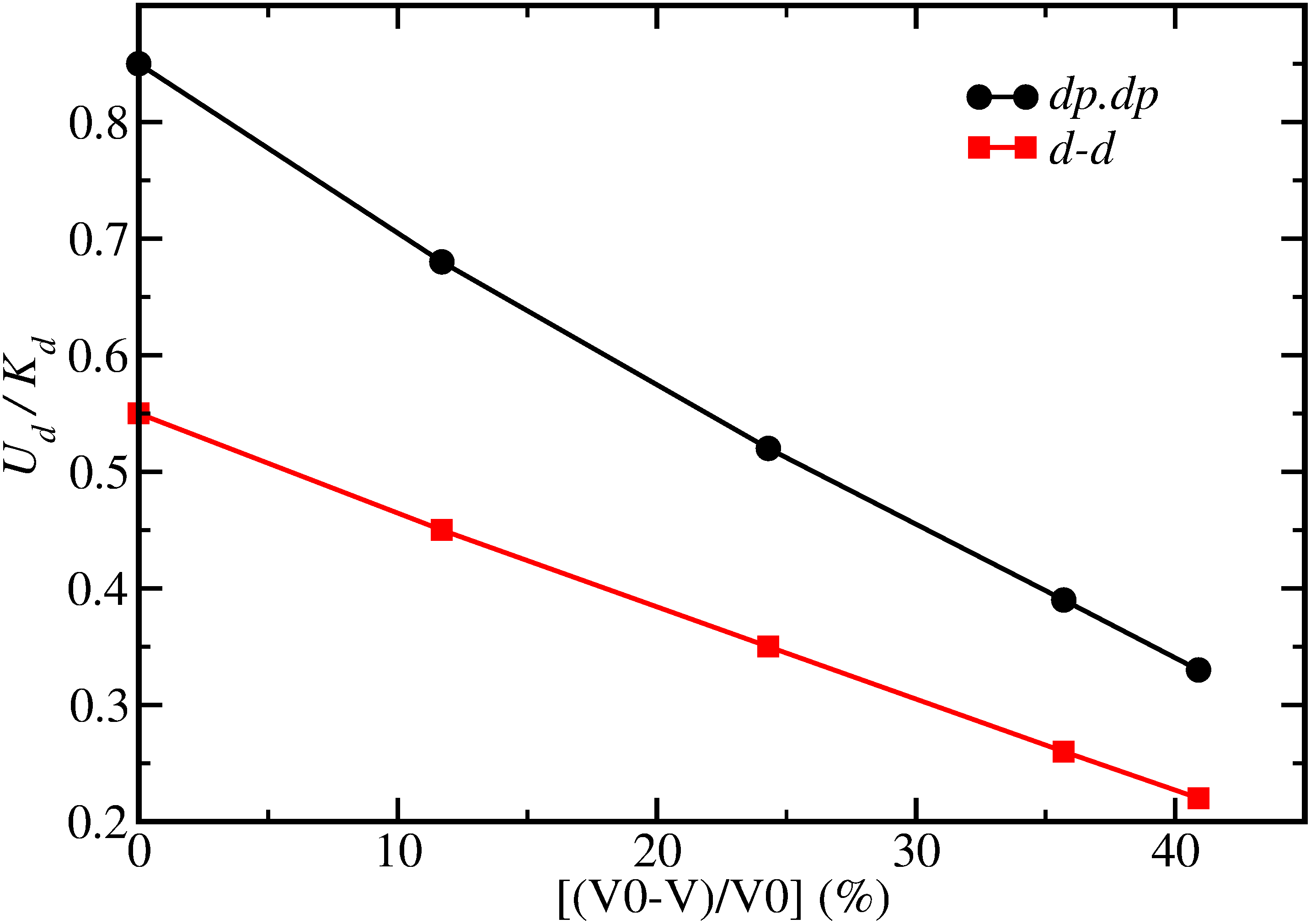}
\caption{Effective correlation strength as a function of compression.}
\label{effInt}
\end{figure}
\begin{figure}
\includegraphics[width=0.99\columnwidth]{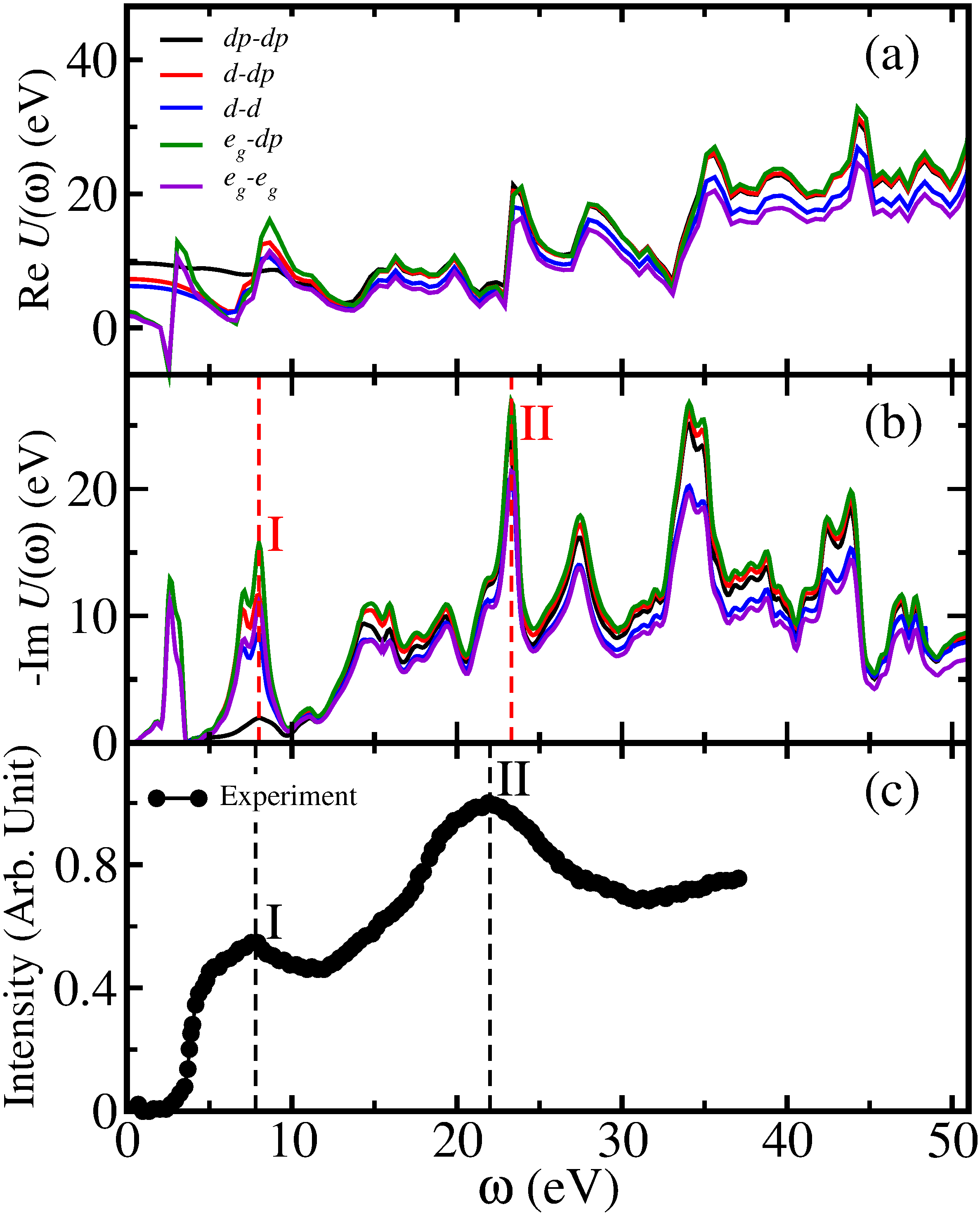}
\caption{Frequency dependence of the (a) real and (b) imaginary part of the partially screened interaction $U(\omega)= \frac{1}{5} \sum_m U_{mm} (\omega)$ for the considered five models with the experimental structure. (b) Experimental electron energy loss spectra are shown to compare the prominent features in $U(\omega)$. Experimental spectra are taken from Ref.~\onlinecite{PhysRevB.49.17293}.}
\label{freqUExp}
\end{figure}
\section{Results and Discussion}
\subsection{Basic electronic structure}
We first analyze the pressure dependence of the basic electronic structure, which is the input to the cRPA.  We looked at the band dispersion and the partial density of states (PDOS), obtained from the non-spin polarized LDA simulation using the experimental lattice parameters. Our results as displayed in Fig.~\ref{bandDOS} show that close to the Fermi level, there are five bands which have predominant Ni-$d$ character. In NiO, Ni is octahedrally coordinated by the O ions. In such a crystal environment Ni-$d$ orbitals split into threefold degenerate $t_{2g}$ states and twofold degenerate $e_g$ states. Since Ni is in $d^8$ configuration, the low lying relatively less dispersed $t_{2g}$ bands are completely filled up, while two $e_{g}$ bands which cross the Fermi level are half filled. Below the Fermi level in the energy range of -8.5 eV to -3.0 eV, we see three bands [blue line in Fig.~\ref{bandDOS}(a)] which are originated from the O-$p$ states as revealed from the PDOS in Fig.~\ref{bandDOS}(b). We also find from Fig.~\ref{bandDOS}(b) that O-$p$ states strongly hybridize with the Ni-$d$ states. As expected Ni-$s$ and $p$ states are located higher in energy in the unoccupied part above the $d$ manifold.
\par
In order to understand the effect of pressure, next we calculated the electronic structure of NiO with a volume of the unit cell that is 59\% of the experimental volume. Our results as displayed in Fig.~\ref{bandDOS1} reveal that pressure induces substantial modification to the electronic structure as expected due to the shrinking of the bond lengths. We clearly see that the Ni-$3d$ and O-$p$ bandwidths grow by a large amount. Such an increase in the bandwidth is related to the reduced interatomic distances at high pressure. The crystal field splitting between $e_g$ and $t_{2g}$ states, and the charge transfer energy (the difference in Ni-$3d$ and O-$2p$ onsite energies) are also found to enhance substantially as a consequence of the shortening of the Ni-O bond lengths. We will see below that such modifications of the electronic structure will have important consequences in modifying the strengths of the screening processes and thus in changing the magnitudes of effective $U$ under the application of pressure. 
\subsection{Pressure induced changes in Coulomb interactions: Static screening}
In this section, we present the bare Coulomb interaction ($v$), and the static (at $\omega$ = 0) values of partially screened ($U$) and fully screened ($W$) Coulomb interaction parameters for a number of unit cell volumes using five different low energy models as mentioned above. We will also analyze effective matrix elements and will check their accuracy with respect to the slater parametrization for the $d$-$d$ and $dp$-$dp$ models.   
\subsubsection{Bare Coulomb interaction ($v$)}
We first focus on the bare values of the interaction parameters as a function of lattice compression. The results of our calculations corresponding to the considered models are displayed in Figs.~\ref{UvW_static}(a)-(e). In all the three models ($dp$-$dp$, $d$-$dp$, $e_g$-$dp$) where an extended $dp$ energy window is considered to define the Wannier function, the values of $v$ for the $t_{2g}$ and $e_g$ states at the experimental lattice volume (V$_0$) come out to be 25.76 eV and 27.08 eV respectively and these values almost remain constant upon variation of the lattice volume [see Figs.~\ref{UvW_static}(a),(b),(d)].  At a highly compressed volume V = 59\% V$_O$, these values become 25.42 eV and 27.53 respectively, implying that a volume compression as high as $\sim$40\%, even hardly changes the localized nature of the Wannier function. An obvious way to change the localization of the Wannier function is to modify the energy window. Within $d$-$d$ model when only $d$ energy window is considered, the magnitudes of $v$ at $V_O$ for the $t_{2g}$ and $e_g$ states reduce to 23.68 eV and 20.93 eV respectively. Interestingly the reduction is much stronger for the $e_g$ states than the $t_{2g}$ states, leading to a smaller value of $v_{e_g}$ than $v_{t_{2g}}$ contrary to the other models where a larger $dp$ energy window is considered. This is expected since a larger energy window allows a stronger hybridization of the $e_g$ states with the O-$p$ states, resulting in a contraction of the spread of the corresponding maximally localized Wannier function and thus giving rise to a larger value of $v_{e_g}$  for the models where the Wannier functions are obtained within a $dp$ energy window. On the other hand due to the symmetry of the $t_{2g}$ states, they could not hybridize strongly with the O-$p$ states as also revealed from our calculated electronic structure (see Fig.~\ref{bandDOS}) and thus the corresponding Wannier functions get less affected due to the modification of the energy window. Coming back to the pressure dependence of $v$ within the $d$-$d$ model, $v_{e_g}$ is again found to remain almost unaltered, however $v_{t_{2g}}$ slightly reduces due to the reduction of the unit-cell volume. The value of $v_{e_g}$ at the experimental volume within $e_g$-$e_g$ model is 20.48 eV which is close to the value obtained in the $d$-$d$ model, implying that the spread of the Wannier functions are similar in these two models. In this model also the bare Coulomb interaction almost does not get affected by the compression of lattice volumes. Therefore we can conclude from these results that the spread of the Wannier functions do not alter significantly under the application of pressure within the limit of our simulations. 
\begin{figure}
\includegraphics[width=0.99\columnwidth]{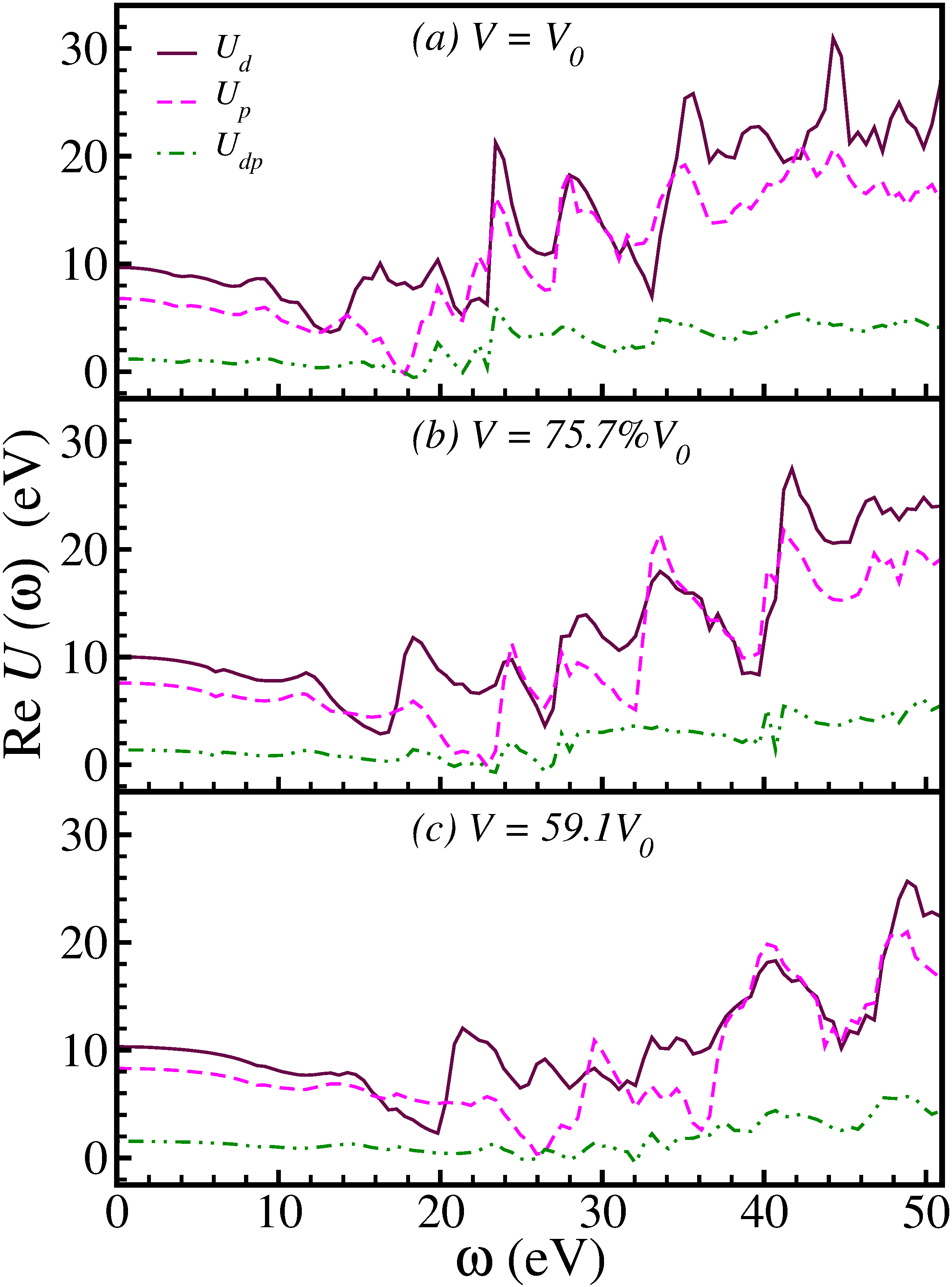}
\caption{Frequency dependence of the partially screened $U_d$, $U_p$, and $U_{dp}$ in the $dp$-$dp$ model for the (a) experimental structure (V = V$_0$), (b) moderately compressed structure (V = 75.7\%V$_0$), and (c) the highest compressed structure in our study (V = 59.1\%V$_0$).}
\label{freqUdp}
\end{figure}
\subsubsection{Partially screened Coulomb interaction ($U$)}
Next we analyze the pressure dependence of the partially screened interaction ($U$), which is the most important parameter needed to construct the low  energy effective model. The results of our calculations are shown in Figs.~\ref{UvW_static}(f)-(j) for all the considered models. We have also tabulated the estimated values of the most relevant parameters corresponding to the experimental structure and the most compressed structure in Table~\ref{UJvalues}. As expected the effective values of $U$ are much reduced than the corresponding $v$ due to the strong screening that arises from the electronic polarizability as discussed before and the strengths of screenings are different for different models as expected. Within the models where $dp$ energy windows were considered, the interactions at each volume gradually decrease as we proceed from $dp$-$dp$, to $d$-$dp$,  and finally to $e_g$-$dp$ model due to the systematic enhancements of the number of screening channels. The values in the $d$-$d$ model are smaller than the corresponding values at the same volume within the $d$-$dp$ model due to the larger extension of Wannier function as evident from the values of bare Coulomb interactions. The same argument is also valid for the $e_g$-$e_g$ model which gives the lowest values of the interaction parameters among all the models. We find that the screened interactions for the $e_g$ states ($U_{e_g}$) are always larger than the corresponding values for $t_{2g}$ states ($U_{t_{2g}}$) at the same volume except the $d$-$d$ model where it show the opposite trend, consistent with the bare interactions.  
\par 
Most importantly our results show that in contrast to the bare interactions, partially screened $U$ increases almost linearly as a function of compression in all the considered models. Such a trend is very much counterintuitive as one would naturally think that pressure will make system more delocalized, reducing the value of effective Coulomb interaction. However this unusual nature of the pressure dependence of effective interaction parameters can be clearly understood by simple analysis of the LDA band structure which is the input to the cRPA calculations. At the level of cRPA, the strengths of electronic screenings are determined by two ingredients: the transition energies and the corresponding matrix elements (overlap integral of the two Wave-functions) [see Eq.~\ref{eq:polarization}]. Our analysis of the electronic structure [see Sec. III(A)] reveal that the compression of lattice volume not only enhances the band width of Ni-$d$, O-$p$ states, and the overlap integrals, but most importantly increases the energy separation between the bonding O-$p$ states and the anti-bonding Ni-$d$ states (transition energies). Since the strengths of different possible screening channels become weaker as the matrix elements reduce and the transition energies enhance, our results of effective $U$ indicate that the later plays the most dominant role, diminishing the polarization as a function of compression. A similar weakening of the screening channels upon compression was reported for the paramagnetic phase of MnO~\cite{PhysRevB.81.115116}. 

It is interesting to mention here that a monotonic growth of the effective $U$ as a function of pressure in $d^8$ ion was also proposed by Ovchinnikov and co-workers~\cite{Ovchinnikov2008,PhysRevB.79.085125} based on a very simple theoretical framework balancing crystal field effects and intra-atomic Hund's exchange. 
However, the physics in that work is quite different, since there the effect stems from the fact that the quantity denoted $U_{eff}$ is in fact the gap and not
strictly speaking the Coulomb parameter. The effect analyzed there is rather related to the analysis of effective interactions (as incorporated the effect of $J$).

We also note that our results of bare and effective interactions at the experimental volume for different models are in good agreement with the previously reported values in Ref.~\onlinecite{PhysRevB.87.165118}. 
\par  
Finally our results suggest that values of $J$ are not that strongly affected by the application of pressure, particularly in the most commonly used $dp$-$dp$, $d$-$dp$, and $d$-$d$ models (see Table~\ref{UJvalues}). We also display the on-site interaction of the $p$ orbitals ($U_{p}$) in Table~\ref{UJvalues} at the experimental volume and  at the smallest volume, calculated  within the $dp$-$dp$ model in which $p$ orbitals are also treated as correlated orbitals alike the $d$ orbitals. We find that this interaction is also substantial and enhanced by 26\% due to the 59\% compression of the unit cell volume. 
Last but not least we note that our estimated value of $U_{p}$ for the experimental structure perfectly agrees with the experimentally reported value of $U_{p}$ in LaCoO$_3$~\cite{PhysRevB.46.9976}. 

\begin{figure}
\includegraphics[width=0.99\columnwidth]{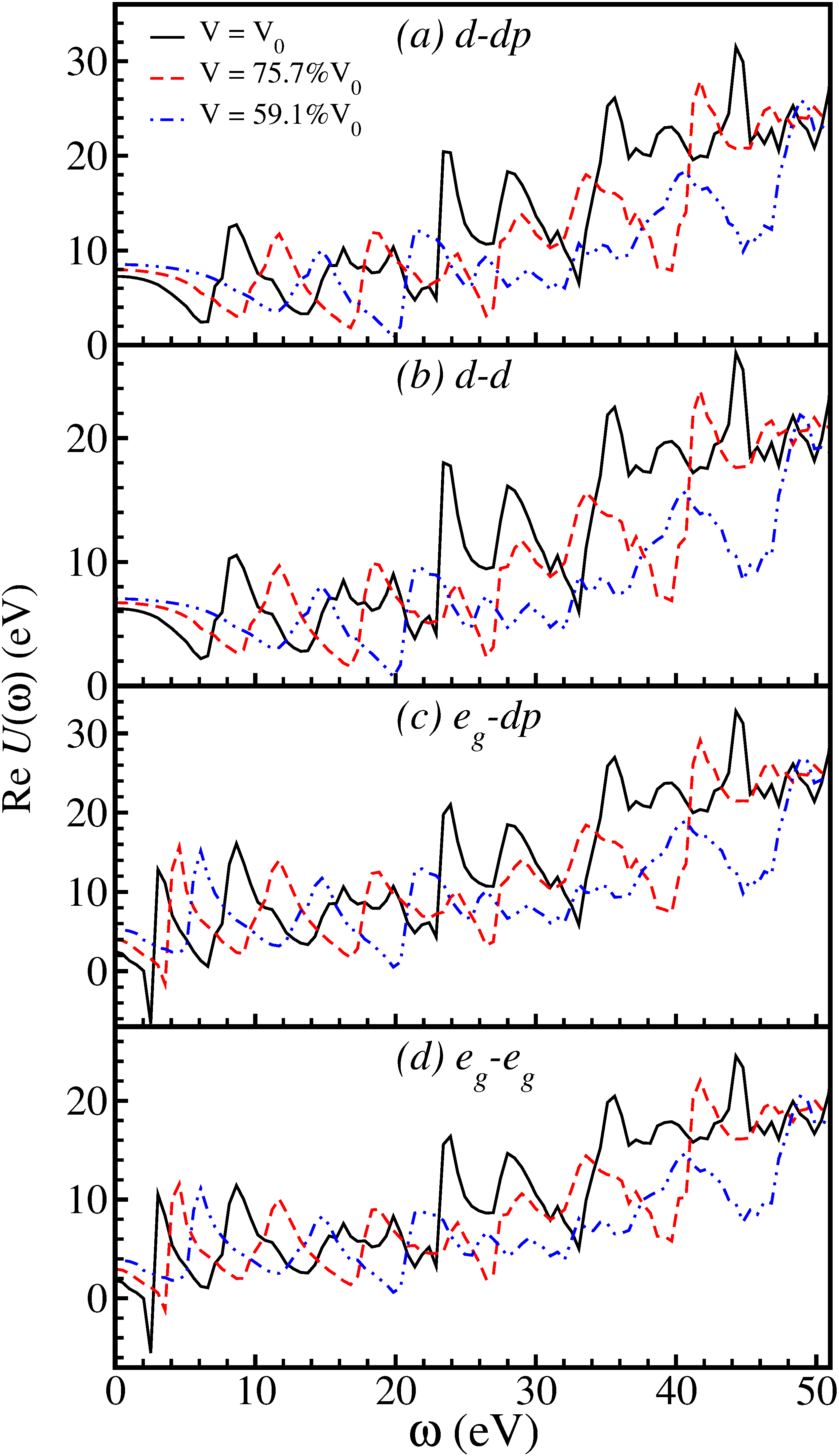}
\caption{Frequency dependence of the partially screened interaction $U(\omega)$ for three different volumes of the unit cell, computed within (a) $d$-$dp$, (b) $d$-$d$, (c) $e_g$-$dp$, and (d) $e_g$-$e_g$ models.}
\label{freqU}
\end{figure}
\bgroup
\def\arraystretch{1.0}
\begin{table}
\caption{Slater parameters at the static limit ($\omega$ = 0) for both the experimental and high pressure structure.}
\vspace{0.1cm}
\begin{tabular}{| c | c c c c c | c c c c c |}
\hline
 & \multicolumn{5}{c|}{V = V$_0$} & \multicolumn{5}{c|}{V = 59.1\%V$_0$} \\
  & \hspace{0.03cm} $F^0$ \hspace{0.03cm} & \hspace{0.03cm} $F^2$ \hspace{0.03cm} & \hspace{0.03cm}  $F^4$ \hspace{0.03cm} & \hspace{0.03cm}  $\frac{F^4}{F^2}$  \hspace{0.03cm} & \hspace{0.03cm} $\frac{F^2+F^4}{14}$ \hspace{0.03cm} & \hspace{0.03cm} $F^0$ \hspace{0.03cm} & \hspace{0.03cm} $F^2$ \hspace{0.03cm} & \hspace{0.03cm}  $F^4$ \hspace{0.03cm} & \hspace{0.03cm}  $\frac{F^4}{F^2}$  \hspace{0.03cm} & \hspace{0.03cm} $\frac{F^2+F^4}{14}$ \hspace{0.03cm} \\[1 ex]
\hline
$dp$-$dp$ & 8.32 & 9.97 & 6.75 & 0.68  & 1.19 & 8.95 & 10.12 & 6.91 & 0.68 & 1.22 \\[1 ex]
$d$-$dp$ & 5.96 & 9.61 & 6.64 & 0.69 & 1.16 & 7.18 & 9.87 & 6.79 & 0.69 & 1.19 \\[1 ex]
$d$-$d$ & 5.13 & 7.86 & 5.50 & 0.70 & 0.95 & 5.96 & 7.67 & 5.25 & 0.68 & 0.92 \\[1 ex]
\hline
\end{tabular}
\label{F0F2F4}
\end{table}
\egroup
\subsubsection{Effective correlation strength ($U/K$)}
The behavior of a correlated system with pressure depends on a delicate balance between the kinetic energy ($K$) and the effective interaction ($U$) of the correlated electrons as a function of pressure.  The relative strength of electron-correlation could be measured by $\tilde{U}$ = $U_d$/$K_d$. We estimate the kinetic energy of the $d$ electrons as 
\begin{equation}
K = \int_{-\infty}^{E_F} \epsilon D(\epsilon)d\epsilon ,
\end{equation}
where $D(\epsilon)$ is the orbital projected density of states of the $d$ orbitals as a function energy and $E_F$ is the Fermi energy. 
We find that $K_d$ increases more strongly than $U_d$ as a function of pressure and as a result $\tilde{U}$ decreases almost linearly with compression for both $d$-$d$ and $dp$-$dp$ models as displayed in Fig~\ref{effInt}. This further confirms that NiO becomes more itinerant under high pressure. 
\subsubsection{Fully screened Coulomb interaction ($W$)}
Next we discuss another extreme case where $v$ is allowed to be screened by all possible channels. As expected the values are further reduced [see Figs.~\ref{UvW_static}(k)-(o)] for all the models. Since all the screening channels are allowed, the interaction parameters for the $t_{2g}$ and $e_{g}$ states are found to be same for all the models namely $dp$-$dp$, $d$-$dp$, and $e_g$-$dp$ models which took the same larger $dp$ energy window to construct the Wannier function. This is simply due to the fact that the bare interactions $v$ are same in magnitude in all these models. Interestingly $W_{t_{2g}}$ is larger than the $W_{e_{g}}$ within the $dp$-$dp$, and $d$-$dp$ models which is opposite to the trend observed for the partially screened parameters. This implies that the internal $e_g$ screening is substantially larger than the internal $t_{2g}$ screening. 
\subsubsection{Inter-site Coulomb interaction: bare and screened}
Finally, we analyze the interatomic bare and partially screened interaction between the nearest neighbor Ni-$d$ orbitals as shown in Table~\ref{Uv_Inter}. We find that both the bare and partially screened interactions are much smaller than the corresponding onsite values as expected. 
The screened interactions $U_{nn}$ increase substantially by lattice compression as was true for the onsite interactions parameters. However we also see a very similar magnitude of increment for the bare interactions, implying that pressure dependence of the intersite $U$ is primarily governed by the Wannier functions, not the screening channels. 
\subsubsection{Accuracy of the Slater parametrization within $dp$-$dp$ and $d$-$d$ models}
We now analyze the accuracy of the Slater parametrization of the interaction matrices for both the experimental structure and the high pressure structure. For the experimental structure within the $dp$-$dp$ model, the partially screened $U$ matrices corresponding to the parallel and opposite spin in the basis of cubic harmonics following the ordering $d_{3z^2-r^2}$, $d_{x^2-y^2}$, $d_{xy}$, $d_{xz}$, and $d_{yz}$ come out to be respectively, 
\begin{align}
U_{mm^\prime}^{\sigma\sigma}\mid_{cRPA} = 
\begin{pmatrix}
   0.00 &  6.80 &  6.60 &  7.75 & 7.75 \\
   6.80 &  0.00 &  8.13 &  6.98 & 6.98 \\
   6.60 &  8.13 &  0.00 &  6.79 & 6.79 \\
   7.75 &  6.98 &  6.79 &  0.00 & 6.79 \\
   7.75 &  6.98 &  6.79 &  6.79 & 0.00 \\
\end{pmatrix}, and\label{eq:umatSameSpindpdpexp}
\end{align}

\begin{align}
U_{mm^\prime}^{\sigma\bar{\sigma}}\mid_{cRPA} = 
\begin{pmatrix}
  10.07 &  7.89 &  7.65 &  8.41 & 8.41 \\
   7.89 & 10.07 &  8.66 &  7.90 & 7.90 \\
   7.65 &  8.66 &  9.41 &  7.67 & 7.67 \\
   8.41 &  7.90 &  7.67 &  9.41 & 7.67 \\
   8.41 &  7.90 &  7.67 &  7.67 & 9.41 \\
\end{pmatrix}\label{eq:umatOpSpindpdpexp}
\end{align}
The above results clearly indicate that there is significant orbital dependence of the intral-orbital interactions. From the diagonal term of the $U_{mm^\prime}^{\sigma\bar{\sigma}}$, we find that the $e_g$ elements are larger by 0.66 eV compared to the $t_{2g}$ elements. Using the estimated values of Slater parameters $F^0$, $F^2$ and $F^4$ as reported in Table~\ref{F0F2F4}, we get the following Slater symmetrized reduced interaction matrices
\begin{align}
U_{mm^\prime}^{\sigma\sigma}\mid_{Slater} = 
\begin{pmatrix}
   0.00 &  6.56 &  6.56 &  7.69 &  7.69 \\
   6.56 &  0.00 &  8.07 &  6.93 &  6.93 \\
   6.56 &  8.07 &  0.00 &  6.93 &  6.93 \\
   7.69 &  6.93 &  6.93 &  0.00 &  6.93 \\
   7.69 &  6.93 &  6.93 &  6.93 &  0.00 \\
\end{pmatrix}, and\label{eq:umatOpSpindpdpexpSlater}
\end{align}

\begin{align}
U_{mm^\prime}^{\sigma\bar{\sigma}}\mid_{Slater} = 
\begin{pmatrix}
   9.69 &  7.60 &  7.60 &  8.36 &  8.36 \\
   7.60 &  9.69 &  8.61 &  7.85 &  7.85 \\
   7.60 &  8.61 &  9.69 &  7.85 &  7.85 \\
   8.36 &  7.85 &  7.85 &  9.69 &  7.85 \\
   8.36 &  7.85 &  7.85 &  7.85 &  9.69 \\
\end{pmatrix}\label{eq:umatSameSpindpdpexpSlater}
\end{align}
We clearly see that the deviation from the directly calculated values are not very large for most of the elements. The largest discrepancy with the direct calculations of around 0.38 eV is observed for the $e_g$ states. The reason of this can be attributed to the stronger hybridization of the $e_g$ states with the O-$p$ states compared to the $t_{2g}$ orbitals, inducing the deviations from the atomic sphericity. Within the Slater parametrization, the intra-orbital interactions $U_{mm}^{\sigma\bar{\sigma}}$ are orbital independent and the magnitude is exactly the same as the average of the diagonal elements of the directly calculated values as displayed in Eqn.~\ref{eq:umatOpSpindpdpexp}. 
The $e_g$-$t_{2g}$ interactions for the opposite spin differ only by about 0.05 eV. 
\par 
The calculated matrices for the highly compressed structure (V = 59.1\%V$_0$) come out to be respectively, 
\begin{align}
U_{mm^\prime}^{\sigma\sigma}\mid_{cRPA} = 
\begin{pmatrix}
   0.00 &  7.55 &  7.22 &  8.39 & 8.39 \\
   7.55 &  0.00 &  8.78 &  7.61 & 7.61 \\
   7.22 &  8.78 &  0.00 &  7.30 & 7.30 \\
   8.39 &  7.61 &  7.30 &  0.00 & 7.30 \\
   8.39 &  7.61 &  7.30 &  7.30 & 0.00 \\
\end{pmatrix}, and\label{eq:umatSameSpindpdpComp}
\end{align}

\begin{align}
U_{mm^\prime}^{\sigma\bar{\sigma}}\mid_{cRPA} = 
\begin{pmatrix}
  10.95 &  8.69 &  8.29 &  9.06 & 9.06 \\
   8.69 & 10.95 &  9.32 &  8.55 & 8.55 \\
   8.29 &  9.32 &  9.91 &  8.17 & 8.17 \\
   9.06 &  8.55 &  8.17 &  9.91 & 8.17 \\
   9.06 &  8.55 &  8.17 &  8.17 & 9.91 \\
\end{pmatrix}\label{eq:umatOpSpindpdpComp}
\end{align}
Our above results find that the orbital dependence become stronger upon compression. The $e_g$ elements are now larger by 1.04 eV compared to the $t_{2g}$ elements. As explained before, a smaller Ni-O distance at high pressure allows a stronger $e_g$-$p$ hybridization, causing a larger value of the corresponding matrix elements. Using the estimated values for the high pressure structure, displayed in Table~\ref{F0F2F4} we again calculate the Slater symmetrized reduced interaction matrix which come out to be 
\begin{align}
U_{mm^\prime}^{\sigma\sigma}\mid_{Slater} = 
\begin{pmatrix}
    0.00 &  7.15 &  7.15 &  8.31 &  8.31 \\
   7.15 &  0.00 &  8.70 &  7.54 &  7.54 \\
   7.15 &  8.70 &  0.00 &  7.54 &  7.54 \\
   8.31 &  7.54 &  7.54 &  0.00 &  7.54 \\
   8.31 &  7.54 &  7.54 &  7.54 &  0.00 \\
\end{pmatrix}, and\label{eq:umatSameSpindpdpCompSlater}
\end{align}

\begin{align}
U_{mm^\prime}^{\sigma\bar{\sigma}}\mid_{Slater} = 
\begin{pmatrix}
  10.34 &  8.22 &  8.22 &  8.99 &  8.99 \\
   8.22 & 10.34 &  9.25 &  8.47 &  8.47 \\
   8.22 &  9.25 & 10.34 &  8.47 &  8.47 \\
   8.99 &  8.47 &  8.47 & 10.34 &  8.47 \\
   8.99 &  8.47 &  8.47 &  8.47 & 10.34 \\
\end{pmatrix}\label{eq:umatOpSpindpdpCompSlater}
\end{align}
Also for the high pressure phase, the differences between the cRPA and the Slater parametrized matrices are not very large for most of the elements. The largest discrepancy of 0.61 eV is again found for the $e_g$ states which is larger by 0.23 eV compared to the experimental structure. A relatively larger deviation from the atomic sphericity results from the stronger covalent nature of the Ni-O bonding at high pressure. 
\par 
Now we will do a very similar analysis for the $d$-$d$ model. Within this model for the experimental structure, the matrices become 
\begin{align}
U_{mm^\prime}^{\sigma\sigma}\mid_{cRPA} = 
\begin{pmatrix}
   0.00 &  3.43 &  3.66 &  4.54 & 4.54 \\
   3.43 &  0.00 &  4.83 &  3.95 & 3.95 \\
   3.66 &  4.83 &  0.00 &  4.27 & 4.27 \\
   4.54 &  3.95 &  4.27 &  0.00 & 4.27 \\
   4.54 &  3.95 &  4.27 &  4.27 & 0.00 \\
\end{pmatrix}, and\label{eq:umatSameSpinddexp}
\end{align}

\begin{align}
U_{mm^\prime}^{\sigma\bar{\sigma}}\mid_{cRPA} = 
\begin{pmatrix}
   5.67 &  4.18 &  4.49 &  5.06 & 5.06 \\
   4.18 &  5.67 &  5.26 &  4.68 & 4.68 \\
   4.49 &  5.26 &  6.61 &  5.06 & 5.06 \\
   5.06 &  4.68 &  5.06 &  6.61 & 5.06 \\
   5.06 &  4.68 &  5.06 &  5.06 & 6.61 \\
\end{pmatrix}\label{eq:umatOpSpinddexp}
\end{align}
These results 
show a similar orbital dependence of $U_{mm}^{\sigma\bar{\sigma}}$ as was observed in $dp$-$dp$ model. For the present case, the $t_{2g}$ elements are found to be larger by 0.94 eV than the $e_g$ elements. 
This is to be contrasted with the opposite finding in the case of the
$dp$-$dp$ calculation. Our interpretation is simple: the larger $e_g$-oxygen
hybridisation leads to more extended Wannier functions for $e_g$ than for
$t_{2g}$ states when only the $d$-bands are included, since the $e_g$-oxygen
hybridisation appears in the form of a ``leakage'' of the $e_g$ Wannier
functions towards the oxygen sites. The inverse is however true when
$d$ and $p$ bands are used for the construction of the Wannier function,
since now the required orthogonalisation confines the stronger hybridising
$e_g$ states more efficiently to their atomic Ni sites.

The Slater symmetrized reduced interaction matrices, obtained by employing the values of Table~\ref{F0F2F4} become 
\begin{align}
U_{mm^\prime}^{\sigma\sigma}\mid_{Slater} = 
\begin{pmatrix}
   0.00 &  3.73 &  3.73 &  4.62 &  4.62 \\
   3.73 &  0.00 &  4.91 &  4.03 &  4.03 \\
   3.73 &  4.91 &  0.00 &  4.03 &  4.03 \\
   4.62 &  4.03 &  4.03 &  0.00 &  4.03 \\
   4.62 &  4.03 &  4.03 &  4.03 &  0.00 \\
\end{pmatrix}, and\label{eq:umatOpSpinddexpSlater}
\end{align}

\begin{align}
U_{mm^\prime}^{\sigma\bar{\sigma}}\mid_{Slater} = 
\begin{pmatrix}
   6.22 &  4.56 &  4.56 &  5.15 &  5.15 \\
   4.56 &  6.22 &  5.35 &  4.76 &  4.76 \\
   4.56 &  5.35 &  6.22 &  4.76 &  4.76 \\
   5.15 &  4.76 &  4.76 &  6.22 &  4.76 \\
   5.15 &  4.76 &  4.76 &  4.76 &  6.22 \\
\end{pmatrix}\label{eq:umatSameSpinddexpSlater}
\end{align}
Again we find that most of the elements are quite close in magnitude to the directly calculated values. The largest discrepancy with the direct calculations of around 0.55 eV is observed for the $e_g$ states. However the averaged directly calculated values coincide with the diagonal elements of the Slater symmetrized matrix. 
\par 
For the high pressure structure the matrices from cRPA calculations become
\begin{align}
U_{mm^\prime}^{\sigma\sigma}\mid_{cRPA} = 
\begin{pmatrix}
   0.00 &  4.42 &  4.57 &  5.44 & 5.44 \\
   4.42 &  0.00 &  5.73 &  4.86 & 4.86 \\
   4.57 &  5.73 &  0.00 &  5.03 & 5.03 \\
   5.44 &  4.86 &  5.03 &  0.00 & 5.03 \\
   5.44 &  4.86 &  5.03 &  5.03 & 0.00 \\
\end{pmatrix}, and\label{eq:umatSameSpinddComp}
\end{align}

\begin{align}
U_{mm^\prime}^{\sigma\bar{\sigma}}\mid_{cRPA} = 
\begin{pmatrix}
   6.73 &  5.19 &  5.36 &  5.94 & 5.94 \\
   5.19 &  6.73 &  6.14 &  5.55 & 5.55 \\
   5.36 &  6.14 &  7.23 &  5.77 & 5.77 \\
   5.94 &  5.55 &  5.77 &  7.23 & 5.77 \\
   5.94 &  5.55 &  5.77 &  5.77 & 7.23 \\
\end{pmatrix}\label{eq:umatOpSpinddComp}
\end{align}
Our above results for the compressed structure find that the orbital dependence of $U_{mm^\prime}^{\sigma\sigma}$ decreases by 0.44 eV with respect to the experimental structure within the $d$-$d$ model, in contrast to the trend obtained in $dp$-$dp$ model. The reason of this has been analyzed before and is attributed to the different screening of those states when $p-d$ screening channels are allowed. The Slater symmetrized reduced interaction matrices for the $d$-$d$ model become
\begin{align}
U_{mm^\prime}^{\sigma\sigma}\mid_{Slater} = 
\begin{pmatrix}
   0.00 &  4.60 &  4.60 &  5.48 &  5.48 \\
   4.60 &  0.00 &  5.77 &  4.89 &  4.89 \\
   4.60 &  5.77 &  0.00 &  4.89 &  4.89 \\
   5.48 &  4.89 &  4.89 &  0.00 &  4.89 \\
   5.48 &  4.89 &  4.89 &  4.89 &  0.00 \\
\end{pmatrix}, and\label{eq:umatOpSpinddCompSlater}
\end{align}

\begin{align}
U_{mm^\prime}^{\sigma\bar{\sigma}}\mid_{Slater} = 
\begin{pmatrix}
   7.01 &  5.40 &  5.40 &  5.99 &  5.99 \\
   5.40 &  7.01 &  6.18 &  5.60 &  5.60 \\
   5.40 &  6.18 &  7.01 &  5.60 &  5.60 \\
   5.99 &  5.60 &  5.60 &  7.01 &  5.60 \\
   5.99 &  5.60 &  5.60 &  5.60 &  7.01 \\
\end{pmatrix}\label{eq:umatSameSpinddCompSlater}
\end{align}
The largest discrepancy with the direct calculations of around 0.28 eV is again found for the $e_g$ states.  Interestingly the deviation reduces substantially, indicating more atomic-like Wannier functions at reduced Ni-O bond distance within $d$-$d$ model.  
\begin{figure}
\includegraphics[width=0.99\columnwidth]{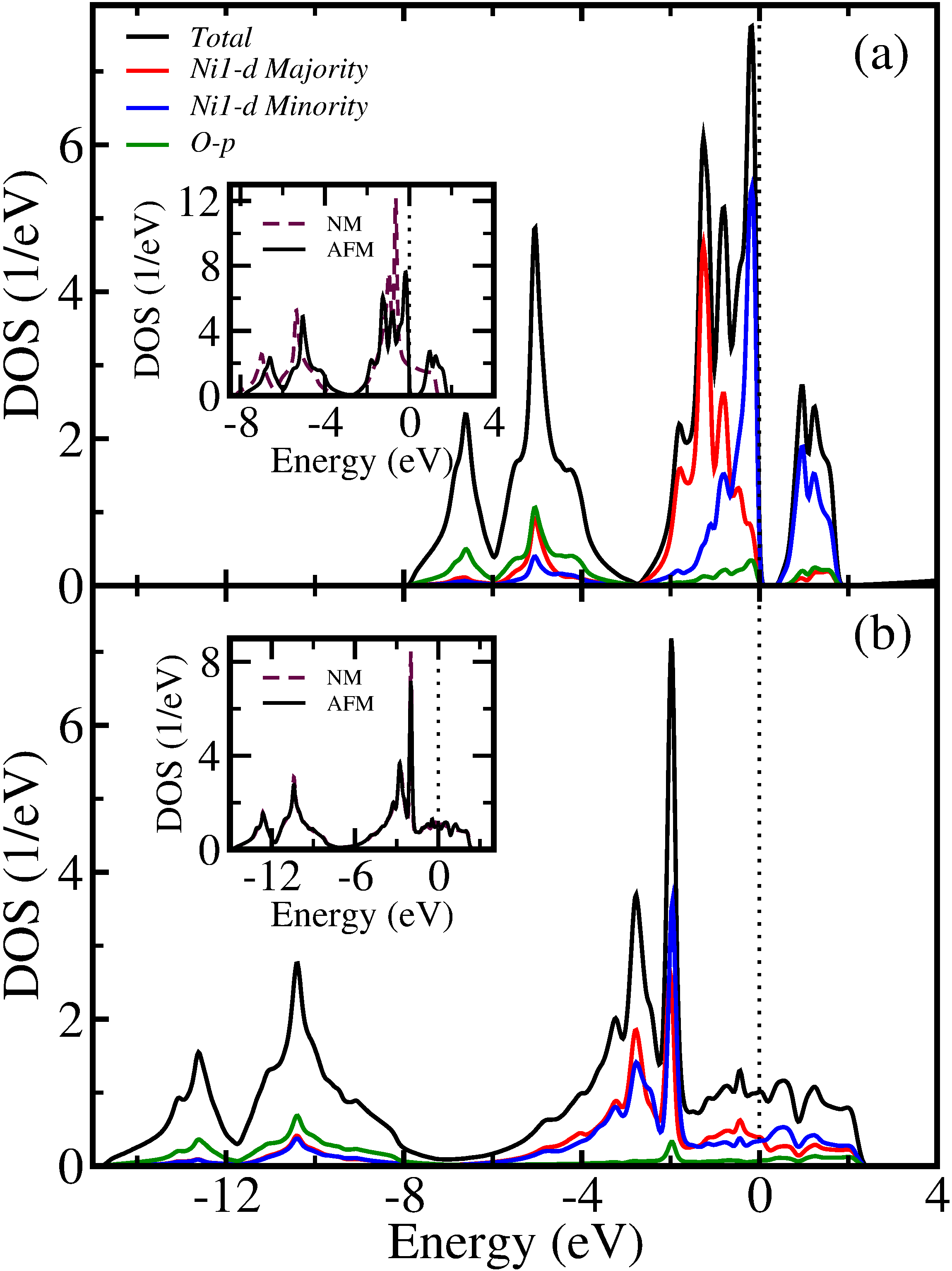}
\caption{Total density of states and site projected partial density of states of Ni-$d$ and O-$p$ orbitals in antiferromagnetic NiO for the (a) experimental structure (V = V$_0$), and (b) the highest compressed structure (V = 59.1\%V$_0$). In the insets, we have compared the total antiferromagnetic DOS with its nonmagnetic counterpart.}
\label{dos_sp}
\end{figure}
\bgroup
\def\arraystretch{1.2}
\begin{table*}
\caption{Averaged bare ($v$) and partially screened ($U$) interactions in antiferromagnetic and non-spin polarized NiO for the two extreme cases (experimental volume V$_0$ and the most compressed volume V = 59\% V$_O$), obtained from $dp$-$dp$ model.}
\vspace{0.1cm}
\begin{tabular}{| c | c c c c c c c | c c c c c c c |}
\hline
 & \multicolumn{7}{c|}{Antiferromagnetic} & \multicolumn{7}{c|}{Non-magnetic} \\
  & \hspace{0.03cm} $v_{d}$ \hspace{0.03cm} & \hspace{0.03cm} $v_{p}$ \hspace{0.03cm}  & \hspace{0.03cm} $v_{dp}$ \hspace{0.03cm} & \hspace{0.03cm}  $U_{d}$ \hspace{0.03cm}  & \hspace{0.03cm} $U_{p}$ \hspace{0.03cm} & \hspace{0.03cm} $U_{dp}$  \hspace{0.03cm}  & \hspace{0.03cm}  $U_d^{SF}$  \hspace{0.03cm} & \hspace{0.03cm} $v_{d}$ \hspace{0.03cm} & \hspace{0.03cm} $v_{p}$ \hspace{0.03cm}  & \hspace{0.03cm} $v_{dp}$ \hspace{0.03cm} & \hspace{0.03cm}  $U_{d}$ \hspace{0.03cm}  & \hspace{0.03cm} $U_{p}$ \hspace{0.03cm} & \hspace{0.03cm} $U_{dp}$  \hspace{0.03cm}  & \hspace{0.03cm}  $U_d^{SF}$  \hspace{0.03cm} \\[1 ex]
\hline
V = V$_0$ &  26.30 & 18.46  & 3.90 & 9.78 & 6.61 & 1.12 & 8.66 & 26.25 & 18.65 & 4.08 & 9.67 & 6.78 & 1.18 & 8.49 \\[1 ex]
V = 59\% V$_O$ & 26.23 & 20.37 & 4.67 & 10.22 & 8.24 & 1.45 & 8.77 & 26.27 & 20.36 & 4.85 & 10.32 & 8.31 & 1.55 & 8.77 \\[1 ex]
\hline
\end{tabular}
\label{Uv_afm}
\end{table*}
\egroup
\subsection{Coulomb interaction: Frequency dependent screening}
In all of the reported LDA+DMFT simulations~\cite{PhysRevB.74.195114,PhysRevLett.99.156404,PhysRevB.75.165115,Karolak201011,PhysRevLett.109.186401,PhysRevB.93.235138,Leonov,
NiO_DMFT1,Nekrasov2012,Kunes2009,PhysRevB.75.165115,PhysRevB.77.195124}, the interactions parameters are considered somewhat different than those calculated within cRPA including the estimated values of this work. It has been found that a large value of $U$ is required to reproduce the experimental band gap as well as to obtain the satellite feature (lower Hubbard band) at the correct binding energy. This fact still remains an unsolved puzzle. In this context the frequency dependence of $U$ may provide a useful insight. With this in mind, in order to understand the necessity of using frequency dependence of $U$ for the experimental and high pressure structures in a realistic many-body model Hamiltonian, we have first compared the computed real and imaginary components of averaged $U(\omega)$ for the experimental structure  among various low energy models in Figs.~\ref{freqUExp}(a) and (b) respectively. 
In the case of $dp$-$dp$ model, the real part of $U(\omega)$ is found to be almost constant in the low frequency region upto around 10 eV. 
This might suggests that the use of the static limit in this model is better justified than in the other setups. This is also expected since no internal transition within the Ni-$3d$ and O-$2p$ states is allowed in this model and the polarization involving the higher lying Ni-$4s$, $4p$ states are responsible for the first prominent peak which appears around 16 eV. However, since this model should of course describe the same
physical situation as in the other models, it is clear that in the $dp$-$dp$ model
the neglect of non-local (and intershell $d$-$p$) interactions is a much more
drastic approximation. Making $U$ frequency-dependent, appears to be 
a way to incorporate nonlocal screening processes in an effective local
description.
Indeed, in all the other models real parts of $U(\omega)$ show a strong variation at low frequencies and correspondingly we also see sharp peaks in their respective imaginary parts, suggesting the failure of static limit. In the $d$-$d$ and $d$-$dp$ model, the first peak in the imaginary parts of $U(\omega)$ appear around 8 eV which is originated from the polarization involving $d$-$p$ transitions. Since the origin of the higher energy peaks in these models are exactly same to the $dp$-$dp$ model, all the other peaks coincide with the peaks of $dp$-$dp$ model as expected. The variation of the real part of $U$ at low $\omega$ become strongest in the $e_g$-$dp$ and $e_g$-$e_g$ models, introducing an additional peak at around 2.6 eV in the imaginary part of $U$. This comes from the polarization involving transition between $t_{2g}$ and $e_g$ states. 
We also observe that a crossover from the low-energy screened regime to the high-energy tail takes place at around 23 eV which can be assigned to a plasma excitation. In order to provide credence to our results of $U(\omega)$, the experimental electron energy loss spectra (EELS), adopted from Ref.~\onlinecite{PhysRevB.49.17293,PhysRevB.69.184404} has been shown in Fig.~\ref{freqUExp}(c). The EELS data of Ref.~\onlinecite{PhysRevB.49.17293} display a well-defined plasmon excitation at 22 eV, providing a very good agreement with our calculated results which is observed around 23 eV (feature II). The other sharp feature (marked as I) at around 8 eV also agrees perfectly with the experimental EELS data. This further validates the use of LDA electronic structure for calculating the effective $U(\omega)$ and most importantly justifies the accuracy of our estimations. We note that in addition to the two sharp characteristic features of EELS data (feature I and II), our computed results also find several other important features at low as well as in high frequency regions. 
\par 
Next we analyze the frequency dependence of $U_d$, $U_p$ and $U_{dp}$ and their pressure dependence within $dp$-$dp$ model in Fig.~\ref{freqUdp}(a)-(c).
We observe that the overall shapes of the $U(\omega)$ are very similar for all the interactions, however the intensities of the peaks are different. This is expected since the number of free electrons involved in the corresponding polarization are different. We observe that compression of the lattice volume does not change the overall nature of the frequency dependence but shift the peak positions towards higher energy. We also notice that the frequency variations of $U_{dp}$ weaken with increasing pressure and almost become constant at the highest pressure of our study. 
\par 
Finally we discuss the pressure dependence of $U(\omega)$ for the other models in Fig.~\ref{freqU}(a)-(d). Again we find that the peak positions are shifted toward higher energy keeping the overall shape of the $U(\omega)$ intact. This can be understood from the evolution of the basic electronic structure due to the application of pressure. As discussed in Sec.III(a), pressure enhances the crystal field splitting between $t_{2g}$ and $e_g$ states as well as the $d$-$p$ energy differences, and therefore the peaks, arising due to the transitions between these states also gradually shift toward higher energy as a function of volume compression. 
\subsection{Coulomb interactions in the spin-polarized phase}
In order to understand the robustness of the computed interaction parameters, we have carried out cRPA calculations for the antiferromagnetic NiO. The computed DOS and projected PDOS of antiferromagnetic phase for the experimental structure and compressed structure are displayed in Fig.~\ref{dos_sp}(a) and (b). As we can see, similar to the non-spin polarized phase, pressure induces an increase in both the $d$ and $p$ bandwidths and their separations. In agreement with the earlier studies, our calculation within LSDA approach also obtained a gaped solution for the experimental structure. However system becomes gapless upon compression. For both the structures, we have also compared the total antiferromagnetic DOS with its nonmagnetic counter part as shown in the inset of Fig.~\ref{dos_sp}(a) and (b). Our results indicate that although there is very little change in the bandwidth upon spin-polarization for the experimental structure, but antiferromagnetic ordering is responsible for opening up the gap at the Fermi level [see inset of Fig.~\ref{dos_sp}(a)].  Owing to the fact that compression also tends to suppress the magnetism, the total antiferromagnetic DOS in the high pressure phase is almost identical to the nonmagnetic total DOS [see Fig.~\ref{dos_sp}(b)]. These features provide an indication that the interaction parameters are likely not to be very different in the antiferromagnetic phase. 

The averaged bare and partially screened interaction parameters are shown in Table~\ref{Uv_afm} for the antiferromagnetic phase as well as the non-spin polarized phase. Interestingly the bare and effective values of $U$ for the $d$ and $p$ orbitals are almost equal in both the phases as was indicated from the analysis of the electronic structure. This further reveals that neither the spreads of the Wannier function, nor the screenings are sensitive to the spin-polarization, establishing the robustness of the quantitative accuracy of our obtained results for the paramagnetic NiO as discussed in this article. 
\section{Conclusions and perspectives}
In summary we have studied the bare and screened Coulomb interactions parameters of NiO under pressure using the cRPA method. Compression induces little changes in the bare Coulomb interactions, indicating only slight modification in the associated Wannier functions, while it gives rise to an enhancement of the effective screened interactions. A detailed analysis of the electronic structure reveals that the compression causes a very large separation between the Ni-$d$ and O-$p$ states, resulting -- at the RPA level -- in a poor screening that leads to this counterintuitive trend of the effective interaction parameters. In contrast, the growth with pressure of the inter-site interactions is primarily governed by a similar enhancement of the corresponding bare interactions. We analyzed the frequency dependence of the effective screened interactions for the experimental equilibrium as well as compressed structures and concluded that it cannot be ignored in a complete description of the electronic structure of NiO. Thus it will be interesting to investigate the renormalization of the Ni-$d$ states within a many-body calculation with frequency dependent $U$, and analyze satellite structures beyond a description with static $U$.
Finally, we have detected a striking similarity in the interaction values between calculations assuming antiferromagnetic and non-magnetic behavior respectively, despite the quite different description of the low-energy states. Photoemission spectroscopy results -- and in particular the close similarity of spectra taken in the antiferromagnetic and paramagnetic phases -- can be interpreted as suggesting that the antiferromagnetic phase might eventually be a more faithful representation even of paramagnetic NiO. In this sense, the agreement of the interaction values in the two calculations is encouraging, questioning the need for further self-consistency.

\subsection{Acknowledgments}
We acknowledge useful discussions with I. Abrikosov, I. Leonov, T. Miyake, and D.D. Sarma.
This work was supported by a Consolidator Grant of the European Research Council
(project CORRELMAT 617196) and by IDRIS/GENCI Orsay under project t2016091393. HJ acknowledges the financial support by the National Natural Science Foundation of China (Projects No. 21373017, 21321001, 21621061).

\begin{thebibliography}{78}%
\makeatletter
\providecommand \@ifxundefined [1]{%
 \@ifx{#1\undefined}
}%
\providecommand \@ifnum [1]{%
 \ifnum #1\expandafter \@firstoftwo
 \else \expandafter \@secondoftwo
 \fi
}%
\providecommand \@ifx [1]{%
 \ifx #1\expandafter \@firstoftwo
 \else \expandafter \@secondoftwo
 \fi
}%
\providecommand \natexlab [1]{#1}%
\providecommand \enquote  [1]{``#1''}%
\providecommand \bibnamefont  [1]{#1}%
\providecommand \bibfnamefont [1]{#1}%
\providecommand \citenamefont [1]{#1}%
\providecommand \href@noop [0]{\@secondoftwo}%
\providecommand \href [0]{\begingroup \@sanitize@url \@href}%
\providecommand \@href[1]{\@@startlink{#1}\@@href}%
\providecommand \@@href[1]{\endgroup#1\@@endlink}%
\providecommand \@sanitize@url [0]{\catcode `\\12\catcode `\$12\catcode
  `\&12\catcode `\#12\catcode `\^12\catcode `\_12\catcode `\%12\relax}%
\providecommand \@@startlink[1]{}%
\providecommand \@@endlink[0]{}%
\providecommand \url  [0]{\begingroup\@sanitize@url \@url }%
\providecommand \@url [1]{\endgroup\@href {#1}{\urlprefix }}%
\providecommand \urlprefix  [0]{URL }%
\providecommand \Eprint [0]{\href }%
\providecommand \doibase [0]{http://dx.doi.org/}%
\providecommand \selectlanguage [0]{\@gobble}%
\providecommand \bibinfo  [0]{\@secondoftwo}%
\providecommand \bibfield  [0]{\@secondoftwo}%
\providecommand \translation [1]{[#1]}%
\providecommand \BibitemOpen [0]{}%
\providecommand \bibitemStop [0]{}%
\providecommand \bibitemNoStop [0]{.\EOS\space}%
\providecommand \EOS [0]{\spacefactor3000\relax}%
\providecommand \BibitemShut  [1]{\csname bibitem#1\endcsname}%
\let\auto@bib@innerbib\@empty
\bibitem [{\citenamefont {Slack}(1960)}]{NiO_AFM_ordering}%
  \BibitemOpen
  \bibfield  {author} {\bibinfo {author} {\bibfnamefont {G.~A.}\ \bibnamefont
  {Slack}},\ }\href {\doibase http://dx.doi.org/10.1063/1.1735895} {\bibfield
  {journal} {\bibinfo  {journal} {Journal of Applied Physics}\ }\textbf
  {\bibinfo {volume} {31}},\ \bibinfo {pages} {1571} (\bibinfo {year}
  {1960})}\BibitemShut {NoStop}%
\bibitem [{\citenamefont {Mott}(1949)}]{Mott_Orig}%
  \BibitemOpen
  \bibfield  {author} {\bibinfo {author} {\bibfnamefont {N.~F.}\ \bibnamefont
  {Mott}},\ }\href@noop {} {\bibfield  {journal} {\bibinfo  {journal}
  {Proceedings of the Physical Society. Section A}\ }\textbf {\bibinfo {volume}
  {62}},\ \bibinfo {pages} {416} (\bibinfo {year} {1949})}\BibitemShut
  {NoStop}%
\bibitem [{\citenamefont {Brandow}(1977)}]{Brandow1977}%
  \BibitemOpen
  \bibfield  {author} {\bibinfo {author} {\bibfnamefont {B.}~\bibnamefont
  {Brandow}},\ }\href {\doibase 10.1080/00018737700101443} {\bibfield
  {journal} {\bibinfo  {journal} {Advances in Physics}\ }\textbf {\bibinfo
  {volume} {26}},\ \bibinfo {pages} {651} (\bibinfo {year} {1977})}\BibitemShut
  {NoStop}%
\bibitem [{\citenamefont {Terakura}\ \emph
  {et~al.}(1984{\natexlab{a}})\citenamefont {Terakura}, \citenamefont
  {Williams}, \citenamefont {Oguchi},\ and\ \citenamefont
  {K\"ubler}}]{PhysRevLett.52.1830}%
  \BibitemOpen
  \bibfield  {author} {\bibinfo {author} {\bibfnamefont {K.}~\bibnamefont
  {Terakura}}, \bibinfo {author} {\bibfnamefont {A.~R.}\ \bibnamefont
  {Williams}}, \bibinfo {author} {\bibfnamefont {T.}~\bibnamefont {Oguchi}}, \
  and\ \bibinfo {author} {\bibfnamefont {J.}~\bibnamefont {K\"ubler}},\ }\href
  {\doibase 10.1103/PhysRevLett.52.1830} {\bibfield  {journal} {\bibinfo
  {journal} {Phys. Rev. Lett.}\ }\textbf {\bibinfo {volume} {52}},\ \bibinfo
  {pages} {1830} (\bibinfo {year} {1984}{\natexlab{a}})}\BibitemShut {NoStop}%
\bibitem [{\citenamefont {Terakura}\ \emph
  {et~al.}(1984{\natexlab{b}})\citenamefont {Terakura}, \citenamefont {Oguchi},
  \citenamefont {Williams},\ and\ \citenamefont {K\"ubler}}]{PhysRevB.30.4734}%
  \BibitemOpen
  \bibfield  {author} {\bibinfo {author} {\bibfnamefont {K.}~\bibnamefont
  {Terakura}}, \bibinfo {author} {\bibfnamefont {T.}~\bibnamefont {Oguchi}},
  \bibinfo {author} {\bibfnamefont {A.~R.}\ \bibnamefont {Williams}}, \ and\
  \bibinfo {author} {\bibfnamefont {J.}~\bibnamefont {K\"ubler}},\ }\href
  {\doibase 10.1103/PhysRevB.30.4734} {\bibfield  {journal} {\bibinfo
  {journal} {Phys. Rev. B}\ }\textbf {\bibinfo {volume} {30}},\ \bibinfo
  {pages} {4734} (\bibinfo {year} {1984}{\natexlab{b}})}\BibitemShut {NoStop}%
\bibitem [{\citenamefont {Eastman}\ and\ \citenamefont
  {Freeouf}(1975)}]{PhysRevLett.34.395}%
  \BibitemOpen
  \bibfield  {author} {\bibinfo {author} {\bibfnamefont {D.~E.}\ \bibnamefont
  {Eastman}}\ and\ \bibinfo {author} {\bibfnamefont {J.~L.}\ \bibnamefont
  {Freeouf}},\ }\href {\doibase 10.1103/PhysRevLett.34.395} {\bibfield
  {journal} {\bibinfo  {journal} {Phys. Rev. Lett.}\ }\textbf {\bibinfo
  {volume} {34}},\ \bibinfo {pages} {395} (\bibinfo {year} {1975})}\BibitemShut
  {NoStop}%
\bibitem [{\citenamefont {Oh}\ \emph {et~al.}(1982)\citenamefont {Oh},
  \citenamefont {Allen}, \citenamefont {Lindau},\ and\ \citenamefont
  {Mikkelsen}}]{PhysRevB.26.4845}%
  \BibitemOpen
  \bibfield  {author} {\bibinfo {author} {\bibfnamefont {S.~J.}\ \bibnamefont
  {Oh}}, \bibinfo {author} {\bibfnamefont {J.~W.}\ \bibnamefont {Allen}},
  \bibinfo {author} {\bibfnamefont {I.}~\bibnamefont {Lindau}}, \ and\ \bibinfo
  {author} {\bibfnamefont {J.~C.}\ \bibnamefont {Mikkelsen}},\ }\href {\doibase
  10.1103/PhysRevB.26.4845} {\bibfield  {journal} {\bibinfo  {journal} {Phys.
  Rev. B}\ }\textbf {\bibinfo {volume} {26}},\ \bibinfo {pages} {4845}
  (\bibinfo {year} {1982})}\BibitemShut {NoStop}%
\bibitem [{\citenamefont {Sawatzky}\ and\ \citenamefont
  {Allen}(1984)}]{PhysRevLett.53.2339}%
  \BibitemOpen
  \bibfield  {author} {\bibinfo {author} {\bibfnamefont {G.~A.}\ \bibnamefont
  {Sawatzky}}\ and\ \bibinfo {author} {\bibfnamefont {J.~W.}\ \bibnamefont
  {Allen}},\ }\href {\doibase 10.1103/PhysRevLett.53.2339} {\bibfield
  {journal} {\bibinfo  {journal} {Phys. Rev. Lett.}\ }\textbf {\bibinfo
  {volume} {53}},\ \bibinfo {pages} {2339} (\bibinfo {year}
  {1984})}\BibitemShut {NoStop}%
\bibitem [{\citenamefont {van~der Laan}\ \emph {et~al.}(1986)\citenamefont
  {van~der Laan}, \citenamefont {Zaanen}, \citenamefont {Sawatzky},
  \citenamefont {Karnatak},\ and\ \citenamefont {Esteva}}]{PhysRevB.33.4253}%
  \BibitemOpen
  \bibfield  {author} {\bibinfo {author} {\bibfnamefont {G.}~\bibnamefont
  {van~der Laan}}, \bibinfo {author} {\bibfnamefont {J.}~\bibnamefont
  {Zaanen}}, \bibinfo {author} {\bibfnamefont {G.~A.}\ \bibnamefont
  {Sawatzky}}, \bibinfo {author} {\bibfnamefont {R.}~\bibnamefont {Karnatak}},
  \ and\ \bibinfo {author} {\bibfnamefont {J.-M.}\ \bibnamefont {Esteva}},\
  }\href {\doibase 10.1103/PhysRevB.33.4253} {\bibfield  {journal} {\bibinfo
  {journal} {Phys. Rev. B}\ }\textbf {\bibinfo {volume} {33}},\ \bibinfo
  {pages} {4253} (\bibinfo {year} {1986})}\BibitemShut {NoStop}%
\bibitem [{\citenamefont {Shen}\ \emph {et~al.}(1991)\citenamefont {Shen},
  \citenamefont {List}, \citenamefont {Dessau}, \citenamefont {Wells},
  \citenamefont {Jepsen}, \citenamefont {Arko}, \citenamefont {Barttlet},
  \citenamefont {Shih}, \citenamefont {Parmigiani}, \citenamefont {Huang},\
  and\ \citenamefont {Lindberg}}]{PhysRevB.44.3604}%
  \BibitemOpen
  \bibfield  {author} {\bibinfo {author} {\bibfnamefont {Z.-X.}\ \bibnamefont
  {Shen}}, \bibinfo {author} {\bibfnamefont {R.~S.}\ \bibnamefont {List}},
  \bibinfo {author} {\bibfnamefont {D.~S.}\ \bibnamefont {Dessau}}, \bibinfo
  {author} {\bibfnamefont {B.~O.}\ \bibnamefont {Wells}}, \bibinfo {author}
  {\bibfnamefont {O.}~\bibnamefont {Jepsen}}, \bibinfo {author} {\bibfnamefont
  {A.~J.}\ \bibnamefont {Arko}}, \bibinfo {author} {\bibfnamefont
  {R.}~\bibnamefont {Barttlet}}, \bibinfo {author} {\bibfnamefont {C.~K.}\
  \bibnamefont {Shih}}, \bibinfo {author} {\bibfnamefont {F.}~\bibnamefont
  {Parmigiani}}, \bibinfo {author} {\bibfnamefont {J.~C.}\ \bibnamefont
  {Huang}}, \ and\ \bibinfo {author} {\bibfnamefont {P.~A.~P.}\ \bibnamefont
  {Lindberg}},\ }\href {\doibase 10.1103/PhysRevB.44.3604} {\bibfield
  {journal} {\bibinfo  {journal} {Phys. Rev. B}\ }\textbf {\bibinfo {volume}
  {44}},\ \bibinfo {pages} {3604} (\bibinfo {year} {1991})}\BibitemShut
  {NoStop}%
\bibitem [{\citenamefont {Tjernberg}\ \emph {et~al.}(1996)\citenamefont
  {Tjernberg}, \citenamefont {S\"oderholm}, \citenamefont {Karlsson},
  \citenamefont {Chiaia}, \citenamefont {Qvarford}, \citenamefont {Nyl\'en},\
  and\ \citenamefont {Lindau}}]{PhysRevB.53.10372}%
  \BibitemOpen
  \bibfield  {author} {\bibinfo {author} {\bibfnamefont {O.}~\bibnamefont
  {Tjernberg}}, \bibinfo {author} {\bibfnamefont {S.}~\bibnamefont
  {S\"oderholm}}, \bibinfo {author} {\bibfnamefont {U.~O.}\ \bibnamefont
  {Karlsson}}, \bibinfo {author} {\bibfnamefont {G.}~\bibnamefont {Chiaia}},
  \bibinfo {author} {\bibfnamefont {M.}~\bibnamefont {Qvarford}}, \bibinfo
  {author} {\bibfnamefont {H.}~\bibnamefont {Nyl\'en}}, \ and\ \bibinfo
  {author} {\bibfnamefont {I.}~\bibnamefont {Lindau}},\ }\href {\doibase
  10.1103/PhysRevB.53.10372} {\bibfield  {journal} {\bibinfo  {journal} {Phys.
  Rev. B}\ }\textbf {\bibinfo {volume} {53}},\ \bibinfo {pages} {10372}
  (\bibinfo {year} {1996})}\BibitemShut {NoStop}%
\bibitem [{\citenamefont {Kune\ifmmode~\check{s}\else \v{s}\fi{}}\ \emph
  {et~al.}(2007{\natexlab{a}})\citenamefont {Kune\ifmmode~\check{s}\else
  \v{s}\fi{}}, \citenamefont {Anisimov}, \citenamefont {Skornyakov},
  \citenamefont {Lukoyanov},\ and\ \citenamefont
  {Vollhardt}}]{PhysRevLett.99.156404}%
  \BibitemOpen
  \bibfield  {author} {\bibinfo {author} {\bibfnamefont {J.}~\bibnamefont
  {Kune\ifmmode~\check{s}\else \v{s}\fi{}}}, \bibinfo {author} {\bibfnamefont
  {V.~I.}\ \bibnamefont {Anisimov}}, \bibinfo {author} {\bibfnamefont {S.~L.}\
  \bibnamefont {Skornyakov}}, \bibinfo {author} {\bibfnamefont {A.~V.}\
  \bibnamefont {Lukoyanov}}, \ and\ \bibinfo {author} {\bibfnamefont
  {D.}~\bibnamefont {Vollhardt}},\ }\href {\doibase
  10.1103/PhysRevLett.99.156404} {\bibfield  {journal} {\bibinfo  {journal}
  {Phys. Rev. Lett.}\ }\textbf {\bibinfo {volume} {99}},\ \bibinfo {pages}
  {156404} (\bibinfo {year} {2007}{\natexlab{a}})}\BibitemShut {NoStop}%
\bibitem [{\citenamefont {Taguchi}\ \emph {et~al.}(2008)\citenamefont
  {Taguchi}, \citenamefont {Matsunami}, \citenamefont {Ishida}, \citenamefont
  {Eguchi}, \citenamefont {Chainani}, \citenamefont {Takata}, \citenamefont
  {Yabashi}, \citenamefont {Tamasaku}, \citenamefont {Nishino}, \citenamefont
  {Ishikawa}, \citenamefont {Senba}, \citenamefont {Ohashi},\ and\
  \citenamefont {Shin}}]{PhysRevLett.100.206401}%
  \BibitemOpen
  \bibfield  {author} {\bibinfo {author} {\bibfnamefont {M.}~\bibnamefont
  {Taguchi}}, \bibinfo {author} {\bibfnamefont {M.}~\bibnamefont {Matsunami}},
  \bibinfo {author} {\bibfnamefont {Y.}~\bibnamefont {Ishida}}, \bibinfo
  {author} {\bibfnamefont {R.}~\bibnamefont {Eguchi}}, \bibinfo {author}
  {\bibfnamefont {A.}~\bibnamefont {Chainani}}, \bibinfo {author}
  {\bibfnamefont {Y.}~\bibnamefont {Takata}}, \bibinfo {author} {\bibfnamefont
  {M.}~\bibnamefont {Yabashi}}, \bibinfo {author} {\bibfnamefont
  {K.}~\bibnamefont {Tamasaku}}, \bibinfo {author} {\bibfnamefont
  {Y.}~\bibnamefont {Nishino}}, \bibinfo {author} {\bibfnamefont
  {T.}~\bibnamefont {Ishikawa}}, \bibinfo {author} {\bibfnamefont
  {Y.}~\bibnamefont {Senba}}, \bibinfo {author} {\bibfnamefont
  {H.}~\bibnamefont {Ohashi}}, \ and\ \bibinfo {author} {\bibfnamefont
  {S.}~\bibnamefont {Shin}},\ }\href {\doibase 10.1103/PhysRevLett.100.206401}
  {\bibfield  {journal} {\bibinfo  {journal} {Phys. Rev. Lett.}\ }\textbf
  {\bibinfo {volume} {100}},\ \bibinfo {pages} {206401} (\bibinfo {year}
  {2008})}\BibitemShut {NoStop}%
\bibitem [{\citenamefont {Weinen}\ \emph {et~al.}(2015)\citenamefont {Weinen},
  \citenamefont {Koethe}, \citenamefont {Chang}, \citenamefont {Agrestini},
  \citenamefont {Kasinathan}, \citenamefont {Liao}, \citenamefont {Fujiwara},
  \citenamefont {Sch{\"{u}}{\ss}ler-Langeheine}, \citenamefont {Strigari},
  \citenamefont {Haupricht}, \citenamefont {Panaccione}, \citenamefont {Offi},
  \citenamefont {Monaco}, \citenamefont {Huotari}, \citenamefont {Tsuei},\ and\
  \citenamefont {Tjeng}}]{Weinen2015}%
  \BibitemOpen
  \bibfield  {author} {\bibinfo {author} {\bibfnamefont {J.}~\bibnamefont
  {Weinen}}, \bibinfo {author} {\bibfnamefont {T.}~\bibnamefont {Koethe}},
  \bibinfo {author} {\bibfnamefont {C.}~\bibnamefont {Chang}}, \bibinfo
  {author} {\bibfnamefont {S.}~\bibnamefont {Agrestini}}, \bibinfo {author}
  {\bibfnamefont {D.}~\bibnamefont {Kasinathan}}, \bibinfo {author}
  {\bibfnamefont {Y.}~\bibnamefont {Liao}}, \bibinfo {author} {\bibfnamefont
  {H.}~\bibnamefont {Fujiwara}}, \bibinfo {author} {\bibfnamefont
  {C.}~\bibnamefont {Sch{\"{u}}{\ss}ler-Langeheine}}, \bibinfo {author}
  {\bibfnamefont {F.}~\bibnamefont {Strigari}}, \bibinfo {author}
  {\bibfnamefont {T.}~\bibnamefont {Haupricht}}, \bibinfo {author}
  {\bibfnamefont {G.}~\bibnamefont {Panaccione}}, \bibinfo {author}
  {\bibfnamefont {F.}~\bibnamefont {Offi}}, \bibinfo {author} {\bibfnamefont
  {G.}~\bibnamefont {Monaco}}, \bibinfo {author} {\bibfnamefont
  {S.}~\bibnamefont {Huotari}}, \bibinfo {author} {\bibfnamefont {K.-D.}\
  \bibnamefont {Tsuei}}, \ and\ \bibinfo {author} {\bibfnamefont
  {L.}~\bibnamefont {Tjeng}},\ }\href {\doibase 10.1016/j.elspec.2014.11.003}
  {\bibfield  {journal} {\bibinfo  {journal} {Journal of Electron Spectroscopy
  and Related Phenomena}\ }\textbf {\bibinfo {volume} {198}},\ \bibinfo {pages}
  {6} (\bibinfo {year} {2015})}\BibitemShut {NoStop}%
\bibitem [{\citenamefont {Fujimori}\ and\ \citenamefont
  {Minami}(1984)}]{PhysRevB.30.957}%
  \BibitemOpen
  \bibfield  {author} {\bibinfo {author} {\bibfnamefont {A.}~\bibnamefont
  {Fujimori}}\ and\ \bibinfo {author} {\bibfnamefont {F.}~\bibnamefont
  {Minami}},\ }\href {\doibase 10.1103/PhysRevB.30.957} {\bibfield  {journal}
  {\bibinfo  {journal} {Phys. Rev. B}\ }\textbf {\bibinfo {volume} {30}},\
  \bibinfo {pages} {957} (\bibinfo {year} {1984})}\BibitemShut {NoStop}%
\bibitem [{\citenamefont {Norman}\ and\ \citenamefont
  {Freeman}(1986)}]{PhysRevB.33.8896}%
  \BibitemOpen
  \bibfield  {author} {\bibinfo {author} {\bibfnamefont {M.~R.}\ \bibnamefont
  {Norman}}\ and\ \bibinfo {author} {\bibfnamefont {A.~J.}\ \bibnamefont
  {Freeman}},\ }\href {\doibase 10.1103/PhysRevB.33.8896} {\bibfield  {journal}
  {\bibinfo  {journal} {Phys. Rev. B}\ }\textbf {\bibinfo {volume} {33}},\
  \bibinfo {pages} {8896} (\bibinfo {year} {1986})}\BibitemShut {NoStop}%
\bibitem [{\citenamefont {Sarma}(1990)}]{Sarma1990}%
  \BibitemOpen
  \bibfield  {author} {\bibinfo {author} {\bibfnamefont {D.~D.}\ \bibnamefont
  {Sarma}},\ }\href@noop {} {\bibfield  {journal} {\bibinfo  {journal} {J. of
  Solid State Chemistry}\ }\textbf {\bibinfo {volume} {88}},\ \bibinfo {pages}
  {45} (\bibinfo {year} {1990})}\BibitemShut {NoStop}%
\bibitem [{\citenamefont {Janssen}\ and\ \citenamefont
  {Nieuwpoort}(1988)}]{PhysRevB.38.3449}%
  \BibitemOpen
  \bibfield  {author} {\bibinfo {author} {\bibfnamefont {G.~J.~M.}\
  \bibnamefont {Janssen}}\ and\ \bibinfo {author} {\bibfnamefont {W.~C.}\
  \bibnamefont {Nieuwpoort}},\ }\href {\doibase 10.1103/PhysRevB.38.3449}
  {\bibfield  {journal} {\bibinfo  {journal} {Phys. Rev. B}\ }\textbf {\bibinfo
  {volume} {38}},\ \bibinfo {pages} {3449} (\bibinfo {year}
  {1988})}\BibitemShut {NoStop}%
\bibitem [{\citenamefont {Shen}\ \emph {et~al.}(1990)\citenamefont {Shen},
  \citenamefont {Shih}, \citenamefont {Jepsen}, \citenamefont {Spicer},
  \citenamefont {Lindau},\ and\ \citenamefont {Allen}}]{PhysRevLett.64.2442}%
  \BibitemOpen
  \bibfield  {author} {\bibinfo {author} {\bibfnamefont {Z.-X.}\ \bibnamefont
  {Shen}}, \bibinfo {author} {\bibfnamefont {C.~K.}\ \bibnamefont {Shih}},
  \bibinfo {author} {\bibfnamefont {O.}~\bibnamefont {Jepsen}}, \bibinfo
  {author} {\bibfnamefont {W.~E.}\ \bibnamefont {Spicer}}, \bibinfo {author}
  {\bibfnamefont {I.}~\bibnamefont {Lindau}}, \ and\ \bibinfo {author}
  {\bibfnamefont {J.~W.}\ \bibnamefont {Allen}},\ }\href {\doibase
  10.1103/PhysRevLett.64.2442} {\bibfield  {journal} {\bibinfo  {journal}
  {Phys. Rev. Lett.}\ }\textbf {\bibinfo {volume} {64}},\ \bibinfo {pages}
  {2442} (\bibinfo {year} {1990})}\BibitemShut {NoStop}%
\bibitem [{\citenamefont {Anisimov}\ \emph {et~al.}(1993)\citenamefont
  {Anisimov}, \citenamefont {Solovyev}, \citenamefont {Korotin}, \citenamefont
  {Czy\ifmmode~\dot{z}\else \.{z}\fi{}yk},\ and\ \citenamefont
  {Sawatzky}}]{PhysRevB.48.16929}%
  \BibitemOpen
  \bibfield  {author} {\bibinfo {author} {\bibfnamefont {V.~I.}\ \bibnamefont
  {Anisimov}}, \bibinfo {author} {\bibfnamefont {I.~V.}\ \bibnamefont
  {Solovyev}}, \bibinfo {author} {\bibfnamefont {M.~A.}\ \bibnamefont
  {Korotin}}, \bibinfo {author} {\bibfnamefont {M.~T.}\ \bibnamefont
  {Czy\ifmmode~\dot{z}\else \.{z}\fi{}yk}}, \ and\ \bibinfo {author}
  {\bibfnamefont {G.~A.}\ \bibnamefont {Sawatzky}},\ }\href {\doibase
  10.1103/PhysRevB.48.16929} {\bibfield  {journal} {\bibinfo  {journal} {Phys.
  Rev. B}\ }\textbf {\bibinfo {volume} {48}},\ \bibinfo {pages} {16929}
  (\bibinfo {year} {1993})}\BibitemShut {NoStop}%
\bibitem [{\citenamefont {Anisimov}\ \emph {et~al.}(1994)\citenamefont
  {Anisimov}, \citenamefont {Kuiper},\ and\ \citenamefont
  {Nordgren}}]{PhysRevB.50.8257}%
  \BibitemOpen
  \bibfield  {author} {\bibinfo {author} {\bibfnamefont {V.~I.}\ \bibnamefont
  {Anisimov}}, \bibinfo {author} {\bibfnamefont {P.}~\bibnamefont {Kuiper}}, \
  and\ \bibinfo {author} {\bibfnamefont {J.}~\bibnamefont {Nordgren}},\ }\href
  {\doibase 10.1103/PhysRevB.50.8257} {\bibfield  {journal} {\bibinfo
  {journal} {Phys. Rev. B}\ }\textbf {\bibinfo {volume} {50}},\ \bibinfo
  {pages} {8257} (\bibinfo {year} {1994})}\BibitemShut {NoStop}%
\bibitem [{\citenamefont {Bengone}\ \emph {et~al.}(2000)\citenamefont
  {Bengone}, \citenamefont {Alouani}, \citenamefont {Bl\"ochl},\ and\
  \citenamefont {Hugel}}]{PhysRevB.62.16392}%
  \BibitemOpen
  \bibfield  {author} {\bibinfo {author} {\bibfnamefont {O.}~\bibnamefont
  {Bengone}}, \bibinfo {author} {\bibfnamefont {M.}~\bibnamefont {Alouani}},
  \bibinfo {author} {\bibfnamefont {P.}~\bibnamefont {Bl\"ochl}}, \ and\
  \bibinfo {author} {\bibfnamefont {J.}~\bibnamefont {Hugel}},\ }\href
  {\doibase 10.1103/PhysRevB.62.16392} {\bibfield  {journal} {\bibinfo
  {journal} {Phys. Rev. B}\ }\textbf {\bibinfo {volume} {62}},\ \bibinfo
  {pages} {16392} (\bibinfo {year} {2000})}\BibitemShut {NoStop}%
\bibitem [{\citenamefont {Ren}\ \emph {et~al.}(2006)\citenamefont {Ren},
  \citenamefont {Leonov}, \citenamefont {Keller}, \citenamefont {Kollar},
  \citenamefont {Nekrasov},\ and\ \citenamefont
  {Vollhardt}}]{PhysRevB.74.195114}%
  \BibitemOpen
  \bibfield  {author} {\bibinfo {author} {\bibfnamefont {X.}~\bibnamefont
  {Ren}}, \bibinfo {author} {\bibfnamefont {I.}~\bibnamefont {Leonov}},
  \bibinfo {author} {\bibfnamefont {G.}~\bibnamefont {Keller}}, \bibinfo
  {author} {\bibfnamefont {M.}~\bibnamefont {Kollar}}, \bibinfo {author}
  {\bibfnamefont {I.}~\bibnamefont {Nekrasov}}, \ and\ \bibinfo {author}
  {\bibfnamefont {D.}~\bibnamefont {Vollhardt}},\ }\href {\doibase
  10.1103/PhysRevB.74.195114} {\bibfield  {journal} {\bibinfo  {journal} {Phys.
  Rev. B}\ }\textbf {\bibinfo {volume} {74}},\ \bibinfo {pages} {195114}
  (\bibinfo {year} {2006})}\BibitemShut {NoStop}%
\bibitem [{\citenamefont {Kotani}\ and\ \citenamefont {van
  Schilfgaarde}(2008)}]{Takao2008}%
  \BibitemOpen
  \bibfield  {author} {\bibinfo {author} {\bibfnamefont {T.}~\bibnamefont
  {Kotani}}\ and\ \bibinfo {author} {\bibfnamefont {M.}~\bibnamefont {van
  Schilfgaarde}},\ }\href {http://stacks.iop.org/0953-8984/20/i=29/a=295214}
  {\bibfield  {journal} {\bibinfo  {journal} {Journal of Physics: Condensed
  Matter}\ }\textbf {\bibinfo {volume} {20}},\ \bibinfo {pages} {295214}
  (\bibinfo {year} {2008})}\BibitemShut {NoStop}%
\bibitem [{\citenamefont {R\"odl}\ \emph {et~al.}(2009)\citenamefont {R\"odl},
  \citenamefont {Fuchs}, \citenamefont {Furthm\"uller},\ and\ \citenamefont
  {Bechstedt}}]{PhysRevB.79.235114}%
  \BibitemOpen
  \bibfield  {author} {\bibinfo {author} {\bibfnamefont {C.}~\bibnamefont
  {R\"odl}}, \bibinfo {author} {\bibfnamefont {F.}~\bibnamefont {Fuchs}},
  \bibinfo {author} {\bibfnamefont {J.}~\bibnamefont {Furthm\"uller}}, \ and\
  \bibinfo {author} {\bibfnamefont {F.}~\bibnamefont {Bechstedt}},\ }\href
  {\doibase 10.1103/PhysRevB.79.235114} {\bibfield  {journal} {\bibinfo
  {journal} {Phys. Rev. B}\ }\textbf {\bibinfo {volume} {79}},\ \bibinfo
  {pages} {235114} (\bibinfo {year} {2009})}\BibitemShut {NoStop}%
\bibitem [{\citenamefont {Thunstr\"om}\ \emph {et~al.}(2012)\citenamefont
  {Thunstr\"om}, \citenamefont {Di~Marco},\ and\ \citenamefont
  {Eriksson}}]{PhysRevLett.109.186401}%
  \BibitemOpen
  \bibfield  {author} {\bibinfo {author} {\bibfnamefont {P.}~\bibnamefont
  {Thunstr\"om}}, \bibinfo {author} {\bibfnamefont {I.}~\bibnamefont
  {Di~Marco}}, \ and\ \bibinfo {author} {\bibfnamefont {O.}~\bibnamefont
  {Eriksson}},\ }\href {\doibase 10.1103/PhysRevLett.109.186401} {\bibfield
  {journal} {\bibinfo  {journal} {Phys. Rev. Lett.}\ }\textbf {\bibinfo
  {volume} {109}},\ \bibinfo {pages} {186401} (\bibinfo {year}
  {2012})}\BibitemShut {NoStop}%
\bibitem [{\citenamefont {Das}\ \emph {et~al.}(2015)\citenamefont {Das},
  \citenamefont {Coulter},\ and\ \citenamefont
  {Manousakis}}]{PhysRevB.91.115105}%
  \BibitemOpen
  \bibfield  {author} {\bibinfo {author} {\bibfnamefont {S.}~\bibnamefont
  {Das}}, \bibinfo {author} {\bibfnamefont {J.~E.}\ \bibnamefont {Coulter}}, \
  and\ \bibinfo {author} {\bibfnamefont {E.}~\bibnamefont {Manousakis}},\
  }\href {\doibase 10.1103/PhysRevB.91.115105} {\bibfield  {journal} {\bibinfo
  {journal} {Phys. Rev. B}\ }\textbf {\bibinfo {volume} {91}},\ \bibinfo
  {pages} {115105} (\bibinfo {year} {2015})}\BibitemShut {NoStop}%
\bibitem [{\citenamefont {Eto}\ \emph {et~al.}(2000)\citenamefont {Eto},
  \citenamefont {Endo}, \citenamefont {Imai}, \citenamefont {Katayama},\ and\
  \citenamefont {Kikegawa}}]{PhysRevB.61.14984}%
  \BibitemOpen
  \bibfield  {author} {\bibinfo {author} {\bibfnamefont {T.}~\bibnamefont
  {Eto}}, \bibinfo {author} {\bibfnamefont {S.}~\bibnamefont {Endo}}, \bibinfo
  {author} {\bibfnamefont {M.}~\bibnamefont {Imai}}, \bibinfo {author}
  {\bibfnamefont {Y.}~\bibnamefont {Katayama}}, \ and\ \bibinfo {author}
  {\bibfnamefont {T.}~\bibnamefont {Kikegawa}},\ }\href {\doibase
  10.1103/PhysRevB.61.14984} {\bibfield  {journal} {\bibinfo  {journal} {Phys.
  Rev. B}\ }\textbf {\bibinfo {volume} {61}},\ \bibinfo {pages} {14984}
  (\bibinfo {year} {2000})}\BibitemShut {NoStop}%
\bibitem [{\citenamefont {Gavrilyuk}\ \emph {et~al.}(2001)\citenamefont
  {Gavrilyuk}, \citenamefont {Troyan}, \citenamefont {Lyubutin},\ and\
  \citenamefont {Sidorov}}]{Gavrilyuk2001}%
  \BibitemOpen
  \bibfield  {author} {\bibinfo {author} {\bibfnamefont {A.~G.}\ \bibnamefont
  {Gavrilyuk}}, \bibinfo {author} {\bibfnamefont {I.~A.}\ \bibnamefont
  {Troyan}}, \bibinfo {author} {\bibfnamefont {I.~S.}\ \bibnamefont
  {Lyubutin}}, \ and\ \bibinfo {author} {\bibfnamefont {V.~A.}\ \bibnamefont
  {Sidorov}},\ }\href {\doibase 10.1134/1.1371350} {\bibfield  {journal}
  {\bibinfo  {journal} {Journal of Experimental and Theoretical Physics}\
  }\textbf {\bibinfo {volume} {92}},\ \bibinfo {pages} {696} (\bibinfo {year}
  {2001})}\BibitemShut {NoStop}%
\bibitem [{\citenamefont {Cohen}\ \emph {et~al.}(1997)\citenamefont {Cohen},
  \citenamefont {Mazin},\ and\ \citenamefont {Isaak}}]{Cohen654}%
  \BibitemOpen
  \bibfield  {author} {\bibinfo {author} {\bibfnamefont {R.~E.}\ \bibnamefont
  {Cohen}}, \bibinfo {author} {\bibfnamefont {I.~I.}\ \bibnamefont {Mazin}}, \
  and\ \bibinfo {author} {\bibfnamefont {D.~G.}\ \bibnamefont {Isaak}},\
  }\href@noop {} {\bibfield  {journal} {\bibinfo  {journal} {Science}\ }\textbf
  {\bibinfo {volume} {275}},\ \bibinfo {pages} {654} (\bibinfo {year}
  {1997})}\BibitemShut {NoStop}%
\bibitem [{\citenamefont {Feng}\ and\ \citenamefont
  {Harrison}(2004)}]{PhysRevB.69.035114}%
  \BibitemOpen
  \bibfield  {author} {\bibinfo {author} {\bibfnamefont {X.-B.}\ \bibnamefont
  {Feng}}\ and\ \bibinfo {author} {\bibfnamefont {N.~M.}\ \bibnamefont
  {Harrison}},\ }\href {\doibase 10.1103/PhysRevB.69.035114} {\bibfield
  {journal} {\bibinfo  {journal} {Phys. Rev. B}\ }\textbf {\bibinfo {volume}
  {69}},\ \bibinfo {pages} {035114} (\bibinfo {year} {2004})}\BibitemShut
  {NoStop}%
\bibitem [{\citenamefont {Gavriliuk}\ \emph {et~al.}(2012)\citenamefont
  {Gavriliuk}, \citenamefont {Trojan},\ and\ \citenamefont
  {Struzhkin}}]{PhysRevLett.109.086402}%
  \BibitemOpen
  \bibfield  {author} {\bibinfo {author} {\bibfnamefont {A.~G.}\ \bibnamefont
  {Gavriliuk}}, \bibinfo {author} {\bibfnamefont {I.~A.}\ \bibnamefont
  {Trojan}}, \ and\ \bibinfo {author} {\bibfnamefont {V.~V.}\ \bibnamefont
  {Struzhkin}},\ }\href {\doibase 10.1103/PhysRevLett.109.086402} {\bibfield
  {journal} {\bibinfo  {journal} {Phys. Rev. Lett.}\ }\textbf {\bibinfo
  {volume} {109}},\ \bibinfo {pages} {086402} (\bibinfo {year}
  {2012})}\BibitemShut {NoStop}%
\bibitem [{\citenamefont {Mott}\ and\ \citenamefont
  {Peierls}(1937)}]{Mott_IMT_NiO}%
  \BibitemOpen
  \bibfield  {author} {\bibinfo {author} {\bibfnamefont {N.~F.}\ \bibnamefont
  {Mott}}\ and\ \bibinfo {author} {\bibfnamefont {R.}~\bibnamefont {Peierls}},\
  }\href@noop {} {\bibfield  {journal} {\bibinfo  {journal} {Proceedings of the
  Physical Society}\ }\textbf {\bibinfo {volume} {49}},\ \bibinfo {pages} {72}
  (\bibinfo {year} {1937})}\BibitemShut {NoStop}%
\bibitem [{\citenamefont {Mattheiss}(1972)}]{PhysRevB.5.290}%
  \BibitemOpen
  \bibfield  {author} {\bibinfo {author} {\bibfnamefont {L.~F.}\ \bibnamefont
  {Mattheiss}},\ }\href {\doibase 10.1103/PhysRevB.5.290} {\bibfield  {journal}
  {\bibinfo  {journal} {Phys. Rev. B}\ }\textbf {\bibinfo {volume} {5}},\
  \bibinfo {pages} {290} (\bibinfo {year} {1972})}\BibitemShut {NoStop}%
\bibitem [{\citenamefont {Cai}\ \emph {et~al.}(2009)\citenamefont {Cai},
  \citenamefont {Han}, \citenamefont {Yu}, \citenamefont {Gao}, \citenamefont
  {Du},\ and\ \citenamefont {Hao}}]{Cai2009}%
  \BibitemOpen
  \bibfield  {author} {\bibinfo {author} {\bibfnamefont {T.}~\bibnamefont
  {Cai}}, \bibinfo {author} {\bibfnamefont {H.}~\bibnamefont {Han}}, \bibinfo
  {author} {\bibfnamefont {Y.}~\bibnamefont {Yu}}, \bibinfo {author}
  {\bibfnamefont {T.}~\bibnamefont {Gao}}, \bibinfo {author} {\bibfnamefont
  {J.}~\bibnamefont {Du}}, \ and\ \bibinfo {author} {\bibfnamefont
  {L.}~\bibnamefont {Hao}},\ }\href {\doibase 10.1016/j.physb.2008.10.009}
  {\bibfield  {journal} {\bibinfo  {journal} {Physica B: Condensed Matter}\
  }\textbf {\bibinfo {volume} {404}},\ \bibinfo {pages} {89} (\bibinfo {year}
  {2009})}\BibitemShut {NoStop}%
\bibitem [{\citenamefont {Hedin}(1965)}]{PhysRev.139.A796}%
  \BibitemOpen
  \bibfield  {author} {\bibinfo {author} {\bibfnamefont {L.}~\bibnamefont
  {Hedin}},\ }\href {\doibase 10.1103/PhysRev.139.A796} {\bibfield  {journal}
  {\bibinfo  {journal} {Phys. Rev.}\ }\textbf {\bibinfo {volume} {139}},\
  \bibinfo {pages} {A796} (\bibinfo {year} {1965})}\BibitemShut {NoStop}%
\bibitem [{\citenamefont {Faleev}\ \emph {et~al.}(2004)\citenamefont {Faleev},
  \citenamefont {van Schilfgaarde},\ and\ \citenamefont
  {Kotani}}]{PhysRevLett.93.126406}%
  \BibitemOpen
  \bibfield  {author} {\bibinfo {author} {\bibfnamefont {S.~V.}\ \bibnamefont
  {Faleev}}, \bibinfo {author} {\bibfnamefont {M.}~\bibnamefont {van
  Schilfgaarde}}, \ and\ \bibinfo {author} {\bibfnamefont {T.}~\bibnamefont
  {Kotani}},\ }\href {\doibase 10.1103/PhysRevLett.93.126406} {\bibfield
  {journal} {\bibinfo  {journal} {Phys. Rev. Lett.}\ }\textbf {\bibinfo
  {volume} {93}},\ \bibinfo {pages} {126406} (\bibinfo {year}
  {2004})}\BibitemShut {NoStop}%
\bibitem [{\citenamefont {Li}\ \emph {et~al.}(2005)\citenamefont {Li},
  \citenamefont {Rignanese},\ and\ \citenamefont {Louie}}]{PhysRevB.71.193102}%
  \BibitemOpen
  \bibfield  {author} {\bibinfo {author} {\bibfnamefont {J.-L.}\ \bibnamefont
  {Li}}, \bibinfo {author} {\bibfnamefont {G.-M.}\ \bibnamefont {Rignanese}}, \
  and\ \bibinfo {author} {\bibfnamefont {S.~G.}\ \bibnamefont {Louie}},\ }\href
  {\doibase 10.1103/PhysRevB.71.193102} {\bibfield  {journal} {\bibinfo
  {journal} {Phys. Rev. B}\ }\textbf {\bibinfo {volume} {71}},\ \bibinfo
  {pages} {193102} (\bibinfo {year} {2005})}\BibitemShut {NoStop}%
\bibitem [{\citenamefont {Jiang}\ \emph {et~al.}(2010)\citenamefont {Jiang},
  \citenamefont {Gomez-Abal}, \citenamefont {Rinke},\ and\ \citenamefont
  {Scheffler}}]{PhysRevB.82.045108}%
  \BibitemOpen
  \bibfield  {author} {\bibinfo {author} {\bibfnamefont {H.}~\bibnamefont
  {Jiang}}, \bibinfo {author} {\bibfnamefont {R.~I.}\ \bibnamefont
  {Gomez-Abal}}, \bibinfo {author} {\bibfnamefont {P.}~\bibnamefont {Rinke}}, \
  and\ \bibinfo {author} {\bibfnamefont {M.}~\bibnamefont {Scheffler}},\ }\href
  {\doibase 10.1103/PhysRevB.82.045108} {\bibfield  {journal} {\bibinfo
  {journal} {Phys. Rev. B}\ }\textbf {\bibinfo {volume} {82}},\ \bibinfo
  {pages} {045108} (\bibinfo {year} {2010})}\BibitemShut {NoStop}%
\bibitem [{Note1()}]{Note1}%
  \BibitemOpen
  \bibinfo {note} {This limitation of GW based approach is also true for other
  transiton metal compounds as shown in Ref.~\protect \rev@citealpnum
  {PhysRevLett.96.226402,PhysRevLett.102.126403}.}\BibitemShut {Stop}%
\bibitem [{\citenamefont {Lichtenstein}\ and\ \citenamefont
  {Katsnelson}(1998)}]{PhysRevB.57.6884}%
  \BibitemOpen
  \bibfield  {author} {\bibinfo {author} {\bibfnamefont {A.~I.}\ \bibnamefont
  {Lichtenstein}}\ and\ \bibinfo {author} {\bibfnamefont {M.~I.}\ \bibnamefont
  {Katsnelson}},\ }\href {\doibase 10.1103/PhysRevB.57.6884} {\bibfield
  {journal} {\bibinfo  {journal} {Phys. Rev. B}\ }\textbf {\bibinfo {volume}
  {57}},\ \bibinfo {pages} {6884} (\bibinfo {year} {1998})}\BibitemShut
  {NoStop}%
\bibitem [{\citenamefont {Georges}\ \emph {et~al.}(1996)\citenamefont
  {Georges}, \citenamefont {Kotliar}, \citenamefont {Krauth},\ and\
  \citenamefont {Rozenberg}}]{RevModPhys.68.13}%
  \BibitemOpen
  \bibfield  {author} {\bibinfo {author} {\bibfnamefont {A.}~\bibnamefont
  {Georges}}, \bibinfo {author} {\bibfnamefont {G.}~\bibnamefont {Kotliar}},
  \bibinfo {author} {\bibfnamefont {W.}~\bibnamefont {Krauth}}, \ and\ \bibinfo
  {author} {\bibfnamefont {M.~J.}\ \bibnamefont {Rozenberg}},\ }\href {\doibase
  10.1103/RevModPhys.68.13} {\bibfield  {journal} {\bibinfo  {journal} {Rev.
  Mod. Phys.}\ }\textbf {\bibinfo {volume} {68}},\ \bibinfo {pages} {13}
  (\bibinfo {year} {1996})}\BibitemShut {NoStop}%
\bibitem [{\citenamefont {Kune\ifmmode~\check{s}\else \v{s}\fi{}}\ \emph
  {et~al.}(2007{\natexlab{b}})\citenamefont {Kune\ifmmode~\check{s}\else
  \v{s}\fi{}}, \citenamefont {Anisimov}, \citenamefont {Lukoyanov},\ and\
  \citenamefont {Vollhardt}}]{PhysRevB.75.165115}%
  \BibitemOpen
  \bibfield  {author} {\bibinfo {author} {\bibfnamefont {J.}~\bibnamefont
  {Kune\ifmmode~\check{s}\else \v{s}\fi{}}}, \bibinfo {author} {\bibfnamefont
  {V.~I.}\ \bibnamefont {Anisimov}}, \bibinfo {author} {\bibfnamefont {A.~V.}\
  \bibnamefont {Lukoyanov}}, \ and\ \bibinfo {author} {\bibfnamefont
  {D.}~\bibnamefont {Vollhardt}},\ }\href {\doibase 10.1103/PhysRevB.75.165115}
  {\bibfield  {journal} {\bibinfo  {journal} {Phys. Rev. B}\ }\textbf {\bibinfo
  {volume} {75}},\ \bibinfo {pages} {165115} (\bibinfo {year}
  {2007}{\natexlab{b}})}\BibitemShut {NoStop}%
\bibitem [{\citenamefont {Karolak}\ \emph {et~al.}(2010)\citenamefont
  {Karolak}, \citenamefont {Ulm}, \citenamefont {Wehling}, \citenamefont
  {Mazurenko}, \citenamefont {Poteryaev},\ and\ \citenamefont
  {Lichtenstein}}]{Karolak201011}%
  \BibitemOpen
  \bibfield  {author} {\bibinfo {author} {\bibfnamefont {M.}~\bibnamefont
  {Karolak}}, \bibinfo {author} {\bibfnamefont {G.}~\bibnamefont {Ulm}},
  \bibinfo {author} {\bibfnamefont {T.}~\bibnamefont {Wehling}}, \bibinfo
  {author} {\bibfnamefont {V.}~\bibnamefont {Mazurenko}}, \bibinfo {author}
  {\bibfnamefont {A.}~\bibnamefont {Poteryaev}}, \ and\ \bibinfo {author}
  {\bibfnamefont {A.}~\bibnamefont {Lichtenstein}},\ }\href {\doibase
  http://dx.doi.org/10.1016/j.elspec.2010.05.021} {\bibfield  {journal}
  {\bibinfo  {journal} {Journal of Electron Spectroscopy and Related
  Phenomena}\ }\textbf {\bibinfo {volume} {181}},\ \bibinfo {pages} {11 }
  (\bibinfo {year} {2010})},\ \bibinfo {note} {proceedings of International
  Workshop on Strong Correlations and Angle-Resolved Photoemission Spectroscopy
  2009}\BibitemShut {NoStop}%
\bibitem [{\citenamefont {Panda}\ \emph {et~al.}(2016)\citenamefont {Panda},
  \citenamefont {Pal}, \citenamefont {Mandal}, \citenamefont {Gorgoi},
  \citenamefont {Das}, \citenamefont {Sarkar}, \citenamefont {Drube},
  \citenamefont {Sun}, \citenamefont {Di~Marco}, \citenamefont {Lindblad},
  \citenamefont {Thunstr\"om}, \citenamefont {Delin}, \citenamefont {Karis},
  \citenamefont {Kvashnin}, \citenamefont {van Schilfgaarde}, \citenamefont
  {Eriksson},\ and\ \citenamefont {Sarma}}]{PhysRevB.93.235138}%
  \BibitemOpen
  \bibfield  {author} {\bibinfo {author} {\bibfnamefont {S.~K.}\ \bibnamefont
  {Panda}}, \bibinfo {author} {\bibfnamefont {B.}~\bibnamefont {Pal}}, \bibinfo
  {author} {\bibfnamefont {S.}~\bibnamefont {Mandal}}, \bibinfo {author}
  {\bibfnamefont {M.}~\bibnamefont {Gorgoi}}, \bibinfo {author} {\bibfnamefont
  {S.}~\bibnamefont {Das}}, \bibinfo {author} {\bibfnamefont {I.}~\bibnamefont
  {Sarkar}}, \bibinfo {author} {\bibfnamefont {W.}~\bibnamefont {Drube}},
  \bibinfo {author} {\bibfnamefont {W.}~\bibnamefont {Sun}}, \bibinfo {author}
  {\bibfnamefont {I.}~\bibnamefont {Di~Marco}}, \bibinfo {author}
  {\bibfnamefont {A.}~\bibnamefont {Lindblad}}, \bibinfo {author}
  {\bibfnamefont {P.}~\bibnamefont {Thunstr\"om}}, \bibinfo {author}
  {\bibfnamefont {A.}~\bibnamefont {Delin}}, \bibinfo {author} {\bibfnamefont
  {O.}~\bibnamefont {Karis}}, \bibinfo {author} {\bibfnamefont {Y.~O.}\
  \bibnamefont {Kvashnin}}, \bibinfo {author} {\bibfnamefont {M.}~\bibnamefont
  {van Schilfgaarde}}, \bibinfo {author} {\bibfnamefont {O.}~\bibnamefont
  {Eriksson}}, \ and\ \bibinfo {author} {\bibfnamefont {D.~D.}\ \bibnamefont
  {Sarma}},\ }\href {\doibase 10.1103/PhysRevB.93.235138} {\bibfield  {journal}
  {\bibinfo  {journal} {Phys. Rev. B}\ }\textbf {\bibinfo {volume} {93}},\
  \bibinfo {pages} {235138} (\bibinfo {year} {2016})}\BibitemShut {NoStop}%
\bibitem [{\citenamefont {Leonov}\ \emph {et~al.}(2016)\citenamefont {Leonov},
  \citenamefont {Pourovskii}, \citenamefont {Georges},\ and\ \citenamefont
  {Abrikosov}}]{Leonov}%
  \BibitemOpen
  \bibfield  {author} {\bibinfo {author} {\bibfnamefont {I.}~\bibnamefont
  {Leonov}}, \bibinfo {author} {\bibfnamefont {L.}~\bibnamefont {Pourovskii}},
  \bibinfo {author} {\bibfnamefont {A.}~\bibnamefont {Georges}}, \ and\
  \bibinfo {author} {\bibfnamefont {I.~A.}\ \bibnamefont {Abrikosov}},\ }\href
  {\doibase 10.1103/PhysRevB.94.155135} {\bibfield  {journal} {\bibinfo
  {journal} {Phys. Rev. B}\ }\textbf {\bibinfo {volume} {94}},\ \bibinfo
  {pages} {155135} (\bibinfo {year} {2016})}\BibitemShut {NoStop}%
\bibitem [{\citenamefont {Skornyakov}\ and\ \citenamefont
  {Anisimov}(2015)}]{NiO_DMFT1}%
  \BibitemOpen
  \bibfield  {author} {\bibinfo {author} {\bibfnamefont {S.}~\bibnamefont
  {Skornyakov}}\ and\ \bibinfo {author} {\bibfnamefont {V.}~\bibnamefont
  {Anisimov}},\ }\href {\doibase 10.1080/14786435.2013.769693} {\bibfield
  {journal} {\bibinfo  {journal} {Philosophical Magazine}\ }\textbf {\bibinfo
  {volume} {95}},\ \bibinfo {pages} {1244} (\bibinfo {year}
  {2015})}\BibitemShut {NoStop}%
\bibitem [{\citenamefont {Nekrasov}\ \emph {et~al.}(2012)\citenamefont
  {Nekrasov}, \citenamefont {Pavlov},\ and\ \citenamefont
  {Sadovskii}}]{Nekrasov2012}%
  \BibitemOpen
  \bibfield  {author} {\bibinfo {author} {\bibfnamefont {I.~A.}\ \bibnamefont
  {Nekrasov}}, \bibinfo {author} {\bibfnamefont {V.~S.}\ \bibnamefont
  {Pavlov}}, \ and\ \bibinfo {author} {\bibfnamefont {M.~V.}\ \bibnamefont
  {Sadovskii}},\ }\href {\doibase 10.1134/S0021364012110070} {\bibfield
  {journal} {\bibinfo  {journal} {JETP Letters}\ }\textbf {\bibinfo {volume}
  {95}},\ \bibinfo {pages} {581} (\bibinfo {year} {2012})}\BibitemShut
  {NoStop}%
\bibitem [{\citenamefont {Kune{\v{s}}}\ \emph {et~al.}(2009)\citenamefont
  {Kune{\v{s}}}, \citenamefont {Leonov}, \citenamefont {Kollar}, \citenamefont
  {Byczuk}, \citenamefont {Anisimov},\ and\ \citenamefont
  {Vollhardt}}]{Kunes2009}%
  \BibitemOpen
  \bibfield  {author} {\bibinfo {author} {\bibfnamefont {J.}~\bibnamefont
  {Kune{\v{s}}}}, \bibinfo {author} {\bibfnamefont {I.}~\bibnamefont {Leonov}},
  \bibinfo {author} {\bibfnamefont {M.}~\bibnamefont {Kollar}}, \bibinfo
  {author} {\bibfnamefont {K.}~\bibnamefont {Byczuk}}, \bibinfo {author}
  {\bibfnamefont {V.~I.}\ \bibnamefont {Anisimov}}, \ and\ \bibinfo {author}
  {\bibfnamefont {D.}~\bibnamefont {Vollhardt}},\ }\href {\doibase
  10.1140/epjst/e2010-01209-0} {\bibfield  {journal} {\bibinfo  {journal} {The
  European Physical Journal Special Topics}\ }\textbf {\bibinfo {volume}
  {180}},\ \bibinfo {pages} {5} (\bibinfo {year} {2009})}\BibitemShut {NoStop}%
\bibitem [{\citenamefont {Miura}\ and\ \citenamefont
  {Fujiwara}(2008)}]{PhysRevB.77.195124}%
  \BibitemOpen
  \bibfield  {author} {\bibinfo {author} {\bibfnamefont {O.}~\bibnamefont
  {Miura}}\ and\ \bibinfo {author} {\bibfnamefont {T.}~\bibnamefont
  {Fujiwara}},\ }\href {\doibase 10.1103/PhysRevB.77.195124} {\bibfield
  {journal} {\bibinfo  {journal} {Phys. Rev. B}\ }\textbf {\bibinfo {volume}
  {77}},\ \bibinfo {pages} {195124} (\bibinfo {year} {2008})}\BibitemShut
  {NoStop}%
\bibitem [{\citenamefont {Aryasetiawan}\ \emph {et~al.}(2004)\citenamefont
  {Aryasetiawan}, \citenamefont {Imada}, \citenamefont {Georges}, \citenamefont
  {Kotliar}, \citenamefont {Biermann},\ and\ \citenamefont
  {Lichtenstein}}]{cRPA_orig}%
  \BibitemOpen
  \bibfield  {author} {\bibinfo {author} {\bibfnamefont {F.}~\bibnamefont
  {Aryasetiawan}}, \bibinfo {author} {\bibfnamefont {M.}~\bibnamefont {Imada}},
  \bibinfo {author} {\bibfnamefont {A.}~\bibnamefont {Georges}}, \bibinfo
  {author} {\bibfnamefont {G.}~\bibnamefont {Kotliar}}, \bibinfo {author}
  {\bibfnamefont {S.}~\bibnamefont {Biermann}}, \ and\ \bibinfo {author}
  {\bibfnamefont {A.~I.}\ \bibnamefont {Lichtenstein}},\ }\href {\doibase
  10.1103/PhysRevB.70.195104} {\bibfield  {journal} {\bibinfo  {journal} {Phys.
  Rev. B}\ }\textbf {\bibinfo {volume} {70}},\ \bibinfo {pages} {195104}
  (\bibinfo {year} {2004})}\BibitemShut {NoStop}%
\bibitem [{\citenamefont {Vaugier}\ \emph {et~al.}(2012)\citenamefont
  {Vaugier}, \citenamefont {Jiang},\ and\ \citenamefont
  {Biermann}}]{cRPA_wien2kImple}%
  \BibitemOpen
  \bibfield  {author} {\bibinfo {author} {\bibfnamefont {L.}~\bibnamefont
  {Vaugier}}, \bibinfo {author} {\bibfnamefont {H.}~\bibnamefont {Jiang}}, \
  and\ \bibinfo {author} {\bibfnamefont {S.}~\bibnamefont {Biermann}},\ }\href
  {\doibase 10.1103/PhysRevB.86.165105} {\bibfield  {journal} {\bibinfo
  {journal} {Phys. Rev. B}\ }\textbf {\bibinfo {volume} {86}},\ \bibinfo
  {pages} {165105} (\bibinfo {year} {2012})}\BibitemShut {NoStop}%
\bibitem [{\citenamefont {Miyake}\ and\ \citenamefont
  {Aryasetiawan}(2008)}]{PhysRevB.77.085122}%
  \BibitemOpen
  \bibfield  {author} {\bibinfo {author} {\bibfnamefont {T.}~\bibnamefont
  {Miyake}}\ and\ \bibinfo {author} {\bibfnamefont {F.}~\bibnamefont
  {Aryasetiawan}},\ }\href {\doibase 10.1103/PhysRevB.77.085122} {\bibfield
  {journal} {\bibinfo  {journal} {Phys. Rev. B}\ }\textbf {\bibinfo {volume}
  {77}},\ \bibinfo {pages} {085122} (\bibinfo {year} {2008})}\BibitemShut
  {NoStop}%
\bibitem [{\citenamefont {Miyake}\ \emph {et~al.}(2009)\citenamefont {Miyake},
  \citenamefont {Aryasetiawan},\ and\ \citenamefont
  {Imada}}]{PhysRevB.80.155134}%
  \BibitemOpen
  \bibfield  {author} {\bibinfo {author} {\bibfnamefont {T.}~\bibnamefont
  {Miyake}}, \bibinfo {author} {\bibfnamefont {F.}~\bibnamefont
  {Aryasetiawan}}, \ and\ \bibinfo {author} {\bibfnamefont {M.}~\bibnamefont
  {Imada}},\ }\href {\doibase 10.1103/PhysRevB.80.155134} {\bibfield  {journal}
  {\bibinfo  {journal} {Phys. Rev. B}\ }\textbf {\bibinfo {volume} {80}},\
  \bibinfo {pages} {155134} (\bibinfo {year} {2009})}\BibitemShut {NoStop}%
\bibitem [{\citenamefont {\ifmmode \mbox{\c{S}}\else \c{S}\fi{}a\ifmmode
  \mbox{\c{s}}\else \c{s}\fi{}\ifmmode \imath \else \i
  \fi{}o\ifmmode~\breve{g}\else \u{g}\fi{}lu}\ \emph
  {et~al.}(2011)\citenamefont {\ifmmode \mbox{\c{S}}\else \c{S}\fi{}a\ifmmode
  \mbox{\c{s}}\else \c{s}\fi{}\ifmmode \imath \else \i
  \fi{}o\ifmmode~\breve{g}\else \u{g}\fi{}lu}, \citenamefont {Friedrich},\ and\
  \citenamefont {Bl\"ugel}}]{PhysRevB.83.121101}%
  \BibitemOpen
  \bibfield  {author} {\bibinfo {author} {\bibfnamefont {E.}~\bibnamefont
  {\ifmmode \mbox{\c{S}}\else \c{S}\fi{}a\ifmmode \mbox{\c{s}}\else
  \c{s}\fi{}\ifmmode \imath \else \i \fi{}o\ifmmode~\breve{g}\else
  \u{g}\fi{}lu}}, \bibinfo {author} {\bibfnamefont {C.}~\bibnamefont
  {Friedrich}}, \ and\ \bibinfo {author} {\bibfnamefont {S.}~\bibnamefont
  {Bl\"ugel}},\ }\href {\doibase 10.1103/PhysRevB.83.121101} {\bibfield
  {journal} {\bibinfo  {journal} {Phys. Rev. B}\ }\textbf {\bibinfo {volume}
  {83}},\ \bibinfo {pages} {121101} (\bibinfo {year} {2011})}\BibitemShut
  {NoStop}%
\bibitem [{\citenamefont {Martins}\ \emph {et~al.}(2011)\citenamefont
  {Martins}, \citenamefont {Aichhorn}, \citenamefont {Vaugier},\ and\
  \citenamefont {Biermann}}]{PhysRevLett.107.266404}%
  \BibitemOpen
  \bibfield  {author} {\bibinfo {author} {\bibfnamefont {C.}~\bibnamefont
  {Martins}}, \bibinfo {author} {\bibfnamefont {M.}~\bibnamefont {Aichhorn}},
  \bibinfo {author} {\bibfnamefont {L.}~\bibnamefont {Vaugier}}, \ and\
  \bibinfo {author} {\bibfnamefont {S.}~\bibnamefont {Biermann}},\ }\href
  {\doibase 10.1103/PhysRevLett.107.266404} {\bibfield  {journal} {\bibinfo
  {journal} {Phys. Rev. Lett.}\ }\textbf {\bibinfo {volume} {107}},\ \bibinfo
  {pages} {266404} (\bibinfo {year} {2011})}\BibitemShut {NoStop}%
\bibitem [{\citenamefont {Nomura}\ \emph {et~al.}(2012)\citenamefont {Nomura},
  \citenamefont {Kaltak}, \citenamefont {Nakamura}, \citenamefont {Taranto},
  \citenamefont {Sakai}, \citenamefont {Toschi}, \citenamefont {Arita},
  \citenamefont {Held}, \citenamefont {Kresse},\ and\ \citenamefont
  {Imada}}]{PhysRevB.86.085117}%
  \BibitemOpen
  \bibfield  {author} {\bibinfo {author} {\bibfnamefont {Y.}~\bibnamefont
  {Nomura}}, \bibinfo {author} {\bibfnamefont {M.}~\bibnamefont {Kaltak}},
  \bibinfo {author} {\bibfnamefont {K.}~\bibnamefont {Nakamura}}, \bibinfo
  {author} {\bibfnamefont {C.}~\bibnamefont {Taranto}}, \bibinfo {author}
  {\bibfnamefont {S.}~\bibnamefont {Sakai}}, \bibinfo {author} {\bibfnamefont
  {A.}~\bibnamefont {Toschi}}, \bibinfo {author} {\bibfnamefont
  {R.}~\bibnamefont {Arita}}, \bibinfo {author} {\bibfnamefont
  {K.}~\bibnamefont {Held}}, \bibinfo {author} {\bibfnamefont {G.}~\bibnamefont
  {Kresse}}, \ and\ \bibinfo {author} {\bibfnamefont {M.}~\bibnamefont
  {Imada}},\ }\href {\doibase 10.1103/PhysRevB.86.085117} {\bibfield  {journal}
  {\bibinfo  {journal} {Phys. Rev. B}\ }\textbf {\bibinfo {volume} {86}},\
  \bibinfo {pages} {085117} (\bibinfo {year} {2012})}\BibitemShut {NoStop}%
\bibitem [{\citenamefont {Shih}\ \emph {et~al.}(2012)\citenamefont {Shih},
  \citenamefont {Abtew}, \citenamefont {Yuan}, \citenamefont {Zhang},\ and\
  \citenamefont {Zhang}}]{PhysRevB.86.165124}%
  \BibitemOpen
  \bibfield  {author} {\bibinfo {author} {\bibfnamefont {B.-C.}\ \bibnamefont
  {Shih}}, \bibinfo {author} {\bibfnamefont {T.~A.}\ \bibnamefont {Abtew}},
  \bibinfo {author} {\bibfnamefont {X.}~\bibnamefont {Yuan}}, \bibinfo {author}
  {\bibfnamefont {W.}~\bibnamefont {Zhang}}, \ and\ \bibinfo {author}
  {\bibfnamefont {P.}~\bibnamefont {Zhang}},\ }\href {\doibase
  10.1103/PhysRevB.86.165124} {\bibfield  {journal} {\bibinfo  {journal} {Phys.
  Rev. B}\ }\textbf {\bibinfo {volume} {86}},\ \bibinfo {pages} {165124}
  (\bibinfo {year} {2012})}\BibitemShut {NoStop}%
\bibitem [{\citenamefont {Panda}\ \emph {et~al.}(2013)\citenamefont {Panda},
  \citenamefont {Dasgupta}, \citenamefont {{\c S}a{\c s}ıo{\u g}lu},
  \citenamefont {Bl{\"u}gel},\ and\ \citenamefont {Sarma}}]{pandaNiS}%
  \BibitemOpen
  \bibfield  {author} {\bibinfo {author} {\bibfnamefont {S.~K.}\ \bibnamefont
  {Panda}}, \bibinfo {author} {\bibfnamefont {I.}~\bibnamefont {Dasgupta}},
  \bibinfo {author} {\bibfnamefont {E.}~\bibnamefont {{\c S}a{\c s}ıo{\u
  g}lu}}, \bibinfo {author} {\bibfnamefont {S.}~\bibnamefont {Bl{\"u}gel}}, \
  and\ \bibinfo {author} {\bibfnamefont {D.~D.}\ \bibnamefont {Sarma}},\
  }\href@noop {} {\bibfield  {journal} {\bibinfo  {journal} {Scientific
  Reports}\ }\textbf {\bibinfo {volume} {3}},\ \bibinfo {pages} {2995}
  (\bibinfo {year} {2013})}\BibitemShut {NoStop}%
\bibitem [{\citenamefont {Amadon}\ \emph {et~al.}(2014)\citenamefont {Amadon},
  \citenamefont {Applencourt},\ and\ \citenamefont
  {Bruneval}}]{PhysRevB.89.125110}%
  \BibitemOpen
  \bibfield  {author} {\bibinfo {author} {\bibfnamefont {B.}~\bibnamefont
  {Amadon}}, \bibinfo {author} {\bibfnamefont {T.}~\bibnamefont {Applencourt}},
  \ and\ \bibinfo {author} {\bibfnamefont {F.}~\bibnamefont {Bruneval}},\
  }\href {\doibase 10.1103/PhysRevB.89.125110} {\bibfield  {journal} {\bibinfo
  {journal} {Phys. Rev. B}\ }\textbf {\bibinfo {volume} {89}},\ \bibinfo
  {pages} {125110} (\bibinfo {year} {2014})}\BibitemShut {NoStop}%
\bibitem [{\citenamefont {van Roekeghem}\ \emph {et~al.}(2016)\citenamefont
  {van Roekeghem}, \citenamefont {Vaugier}, \citenamefont {Jiang},\ and\
  \citenamefont {Biermann}}]{PhysRevB.94.125147}%
  \BibitemOpen
  \bibfield  {author} {\bibinfo {author} {\bibfnamefont {A.}~\bibnamefont {van
  Roekeghem}}, \bibinfo {author} {\bibfnamefont {L.}~\bibnamefont {Vaugier}},
  \bibinfo {author} {\bibfnamefont {H.}~\bibnamefont {Jiang}}, \ and\ \bibinfo
  {author} {\bibfnamefont {S.}~\bibnamefont {Biermann}},\ }\href {\doibase
  10.1103/PhysRevB.94.125147} {\bibfield  {journal} {\bibinfo  {journal} {Phys.
  Rev. B}\ }\textbf {\bibinfo {volume} {94}},\ \bibinfo {pages} {125147}
  (\bibinfo {year} {2016})}\BibitemShut {NoStop}%
\bibitem [{\citenamefont {Tomczak}\ \emph {et~al.}(2009)\citenamefont
  {Tomczak}, \citenamefont {Miyake}, \citenamefont {Sakuma},\ and\
  \citenamefont {Aryasetiawan}}]{PhysRevB.79.235133}%
  \BibitemOpen
  \bibfield  {author} {\bibinfo {author} {\bibfnamefont {J.~M.}\ \bibnamefont
  {Tomczak}}, \bibinfo {author} {\bibfnamefont {T.}~\bibnamefont {Miyake}},
  \bibinfo {author} {\bibfnamefont {R.}~\bibnamefont {Sakuma}}, \ and\ \bibinfo
  {author} {\bibfnamefont {F.}~\bibnamefont {Aryasetiawan}},\ }\href {\doibase
  10.1103/PhysRevB.79.235133} {\bibfield  {journal} {\bibinfo  {journal} {Phys.
  Rev. B}\ }\textbf {\bibinfo {volume} {79}},\ \bibinfo {pages} {235133}
  (\bibinfo {year} {2009})}\BibitemShut {NoStop}%
\bibitem [{\citenamefont {Tomczak}\ \emph {et~al.}(2010)\citenamefont
  {Tomczak}, \citenamefont {Miyake},\ and\ \citenamefont
  {Aryasetiawan}}]{PhysRevB.81.115116}%
  \BibitemOpen
  \bibfield  {author} {\bibinfo {author} {\bibfnamefont {J.~M.}\ \bibnamefont
  {Tomczak}}, \bibinfo {author} {\bibfnamefont {T.}~\bibnamefont {Miyake}}, \
  and\ \bibinfo {author} {\bibfnamefont {F.}~\bibnamefont {Aryasetiawan}},\
  }\href {\doibase 10.1103/PhysRevB.81.115116} {\bibfield  {journal} {\bibinfo
  {journal} {Phys. Rev. B}\ }\textbf {\bibinfo {volume} {81}},\ \bibinfo
  {pages} {115116} (\bibinfo {year} {2010})}\BibitemShut {NoStop}%
\bibitem [{\citenamefont {Biermann}(2014)}]{0953-8984-26-17-173202}%
  \BibitemOpen
  \bibfield  {author} {\bibinfo {author} {\bibfnamefont {S.}~\bibnamefont
  {Biermann}},\ }\href@noop {} {\bibfield  {journal} {\bibinfo  {journal}
  {Journal of Physics: Condensed Matter}\ }\textbf {\bibinfo {volume} {26}},\
  \bibinfo {pages} {173202} (\bibinfo {year} {2014})}\BibitemShut {NoStop}%
\bibitem [{\citenamefont {Seth}\ \emph {et~al.}()\citenamefont {Seth},
  \citenamefont {Hansmann}, \citenamefont {Roekeghem}, \citenamefont
  {Vaugier},\ and\ \citenamefont {Biermann}}]{Seth-et-al}%
  \BibitemOpen
  \bibfield  {author} {\bibinfo {author} {\bibfnamefont {P.}~\bibnamefont
  {Seth}}, \bibinfo {author} {\bibfnamefont {P.}~\bibnamefont {Hansmann}},
  \bibinfo {author} {\bibfnamefont {A.~v.}\ \bibnamefont {Roekeghem}}, \bibinfo
  {author} {\bibfnamefont {L.}~\bibnamefont {Vaugier}}, \ and\ \bibinfo
  {author} {\bibfnamefont {S.}~\bibnamefont {Biermann}},\ }\href@noop {} {\
  }\Eprint {http://arxiv.org/abs/arXiv:1508.07466v1} {arXiv:1508.07466v1}
  \BibitemShut {NoStop}%
\bibitem [{\citenamefont {Sakuma}\ and\ \citenamefont
  {Aryasetiawan}(2013)}]{PhysRevB.87.165118}%
  \BibitemOpen
  \bibfield  {author} {\bibinfo {author} {\bibfnamefont {R.}~\bibnamefont
  {Sakuma}}\ and\ \bibinfo {author} {\bibfnamefont {F.}~\bibnamefont
  {Aryasetiawan}},\ }\href {\doibase 10.1103/PhysRevB.87.165118} {\bibfield
  {journal} {\bibinfo  {journal} {Phys. Rev. B}\ }\textbf {\bibinfo {volume}
  {87}},\ \bibinfo {pages} {165118} (\bibinfo {year} {2013})}\BibitemShut
  {NoStop}%
\bibitem [{\citenamefont {Kunes}\ \emph {et~al.}(2008)\citenamefont {Kunes},
  \citenamefont {Lukoyanov}, \citenamefont {Anisimov}, \citenamefont
  {Scalettar},\ and\ \citenamefont {Pickett}}]{KunesMnO}%
  \BibitemOpen
  \bibfield  {author} {\bibinfo {author} {\bibfnamefont {J.}~\bibnamefont
  {Kunes}}, \bibinfo {author} {\bibfnamefont {A.~V.}\ \bibnamefont
  {Lukoyanov}}, \bibinfo {author} {\bibfnamefont {V.~I.}\ \bibnamefont
  {Anisimov}}, \bibinfo {author} {\bibfnamefont {R.~T.}\ \bibnamefont
  {Scalettar}}, \ and\ \bibinfo {author} {\bibfnamefont {W.~E.}\ \bibnamefont
  {Pickett}},\ }\href@noop {} {\bibfield  {journal} {\bibinfo  {journal} {Nat
  Mater}\ }\textbf {\bibinfo {volume} {7}},\ \bibinfo {pages} {198} (\bibinfo
  {year} {2008})}\BibitemShut {NoStop}%
\bibitem [{\citenamefont {Ovchinnikov}(2008)}]{Ovchinnikov2008}%
  \BibitemOpen
  \bibfield  {author} {\bibinfo {author} {\bibfnamefont {S.~G.}\ \bibnamefont
  {Ovchinnikov}},\ }\href {\doibase 10.1134/S1063776108070145} {\bibfield
  {journal} {\bibinfo  {journal} {Journal of Experimental and Theoretical
  Physics}\ }\textbf {\bibinfo {volume} {107}},\ \bibinfo {pages} {140}
  (\bibinfo {year} {2008})}\BibitemShut {NoStop}%
\bibitem [{\citenamefont {Lyubutin}\ \emph {et~al.}(2009)\citenamefont
  {Lyubutin}, \citenamefont {Ovchinnikov}, \citenamefont {Gavriliuk},\ and\
  \citenamefont {Struzhkin}}]{PhysRevB.79.085125}%
  \BibitemOpen
  \bibfield  {author} {\bibinfo {author} {\bibfnamefont {I.~S.}\ \bibnamefont
  {Lyubutin}}, \bibinfo {author} {\bibfnamefont {S.~G.}\ \bibnamefont
  {Ovchinnikov}}, \bibinfo {author} {\bibfnamefont {A.~G.}\ \bibnamefont
  {Gavriliuk}}, \ and\ \bibinfo {author} {\bibfnamefont {V.~V.}\ \bibnamefont
  {Struzhkin}},\ }\href {\doibase 10.1103/PhysRevB.79.085125} {\bibfield
  {journal} {\bibinfo  {journal} {Phys. Rev. B}\ }\textbf {\bibinfo {volume}
  {79}},\ \bibinfo {pages} {085125} (\bibinfo {year} {2009})}\BibitemShut
  {NoStop}%
\bibitem [{\citenamefont {Sjöstedt}\ \emph {et~al.}(2000)\citenamefont
  {Sjöstedt}, \citenamefont {Nordström},\ and\ \citenamefont
  {Singh}}]{lapwPluslo}%
  \BibitemOpen
  \bibfield  {author} {\bibinfo {author} {\bibfnamefont {E.}~\bibnamefont
  {Sjöstedt}}, \bibinfo {author} {\bibfnamefont {L.}~\bibnamefont
  {Nordström}}, \ and\ \bibinfo {author} {\bibfnamefont {D.}~\bibnamefont
  {Singh}},\ }\href {\doibase http://dx.doi.org/10.1016/S0038-1098(99)00577-3}
  {\bibfield  {journal} {\bibinfo  {journal} {Solid State Communications}\
  }\textbf {\bibinfo {volume} {114}},\ \bibinfo {pages} {15 } (\bibinfo {year}
  {2000})}\BibitemShut {NoStop}%
\bibitem [{\citenamefont {Schwarz}\ and\ \citenamefont {Blaha}(2003)}]{wien2k}%
  \BibitemOpen
  \bibfield  {author} {\bibinfo {author} {\bibfnamefont {K.}~\bibnamefont
  {Schwarz}}\ and\ \bibinfo {author} {\bibfnamefont {P.}~\bibnamefont
  {Blaha}},\ }\href {\doibase http://dx.doi.org/10.1016/S0927-0256(03)00112-5}
  {\bibfield  {journal} {\bibinfo  {journal} {Computational Materials Science}\
  }\textbf {\bibinfo {volume} {28}},\ \bibinfo {pages} {259 } (\bibinfo {year}
  {2003})}\BibitemShut {NoStop}%
\bibitem [{\citenamefont {Hohenberg}\ and\ \citenamefont {Kohn}(1964)}]{DFT1}%
  \BibitemOpen
  \bibfield  {author} {\bibinfo {author} {\bibfnamefont {P.}~\bibnamefont
  {Hohenberg}}\ and\ \bibinfo {author} {\bibfnamefont {W.}~\bibnamefont
  {Kohn}},\ }\href {\doibase 10.1103/PhysRev.136.B864} {\bibfield  {journal}
  {\bibinfo  {journal} {Phys. Rev.}\ }\textbf {\bibinfo {volume} {136}},\
  \bibinfo {pages} {B864} (\bibinfo {year} {1964})}\BibitemShut {NoStop}%
\bibitem [{\citenamefont {Jones}\ and\ \citenamefont
  {Gunnarsson}(1989)}]{DFT2}%
  \BibitemOpen
  \bibfield  {author} {\bibinfo {author} {\bibfnamefont {R.~O.}\ \bibnamefont
  {Jones}}\ and\ \bibinfo {author} {\bibfnamefont {O.}~\bibnamefont
  {Gunnarsson}},\ }\href {\doibase 10.1103/RevModPhys.61.689} {\bibfield
  {journal} {\bibinfo  {journal} {Rev. Mod. Phys.}\ }\textbf {\bibinfo {volume}
  {61}},\ \bibinfo {pages} {689} (\bibinfo {year} {1989})}\BibitemShut
  {NoStop}%
\bibitem [{\citenamefont {Gorschl\"uter}\ and\ \citenamefont
  {Merz}(1994)}]{PhysRevB.49.17293}%
  \BibitemOpen
  \bibfield  {author} {\bibinfo {author} {\bibfnamefont {A.}~\bibnamefont
  {Gorschl\"uter}}\ and\ \bibinfo {author} {\bibfnamefont {H.}~\bibnamefont
  {Merz}},\ }\href {\doibase 10.1103/PhysRevB.49.17293} {\bibfield  {journal}
  {\bibinfo  {journal} {Phys. Rev. B}\ }\textbf {\bibinfo {volume} {49}},\
  \bibinfo {pages} {17293} (\bibinfo {year} {1994})}\BibitemShut {NoStop}%
\bibitem [{\citenamefont {Chainani}\ \emph {et~al.}(1992)\citenamefont
  {Chainani}, \citenamefont {Mathew},\ and\ \citenamefont
  {Sarma}}]{PhysRevB.46.9976}%
  \BibitemOpen
  \bibfield  {author} {\bibinfo {author} {\bibfnamefont {A.}~\bibnamefont
  {Chainani}}, \bibinfo {author} {\bibfnamefont {M.}~\bibnamefont {Mathew}}, \
  and\ \bibinfo {author} {\bibfnamefont {D.~D.}\ \bibnamefont {Sarma}},\ }\href
  {\doibase 10.1103/PhysRevB.46.9976} {\bibfield  {journal} {\bibinfo
  {journal} {Phys. Rev. B}\ }\textbf {\bibinfo {volume} {46}},\ \bibinfo
  {pages} {9976} (\bibinfo {year} {1992})}\BibitemShut {NoStop}%
\bibitem [{\citenamefont {Dobysheva}\ \emph {et~al.}(2004)\citenamefont
  {Dobysheva}, \citenamefont {Potapov},\ and\ \citenamefont
  {Schryvers}}]{PhysRevB.69.184404}%
  \BibitemOpen
  \bibfield  {author} {\bibinfo {author} {\bibfnamefont {L.~V.}\ \bibnamefont
  {Dobysheva}}, \bibinfo {author} {\bibfnamefont {P.~L.}\ \bibnamefont
  {Potapov}}, \ and\ \bibinfo {author} {\bibfnamefont {D.}~\bibnamefont
  {Schryvers}},\ }\href {\doibase 10.1103/PhysRevB.69.184404} {\bibfield
  {journal} {\bibinfo  {journal} {Phys. Rev. B}\ }\textbf {\bibinfo {volume}
  {69}},\ \bibinfo {pages} {184404} (\bibinfo {year} {2004})}\BibitemShut
  {NoStop}%
\bibitem [{\citenamefont {van Schilfgaarde}\ \emph {et~al.}(2006)\citenamefont
  {van Schilfgaarde}, \citenamefont {Kotani},\ and\ \citenamefont
  {Faleev}}]{PhysRevLett.96.226402}%
  \BibitemOpen
  \bibfield  {author} {\bibinfo {author} {\bibfnamefont {M.}~\bibnamefont {van
  Schilfgaarde}}, \bibinfo {author} {\bibfnamefont {T.}~\bibnamefont {Kotani}},
  \ and\ \bibinfo {author} {\bibfnamefont {S.}~\bibnamefont {Faleev}},\ }\href
  {\doibase 10.1103/PhysRevLett.96.226402} {\bibfield  {journal} {\bibinfo
  {journal} {Phys. Rev. Lett.}\ }\textbf {\bibinfo {volume} {96}},\ \bibinfo
  {pages} {226402} (\bibinfo {year} {2006})}\BibitemShut {NoStop}%
\bibitem [{\citenamefont {Jiang}\ \emph {et~al.}(2009)\citenamefont {Jiang},
  \citenamefont {Gomez-Abal}, \citenamefont {Rinke},\ and\ \citenamefont
  {Scheffler}}]{PhysRevLett.102.126403}%
  \BibitemOpen
  \bibfield  {author} {\bibinfo {author} {\bibfnamefont {H.}~\bibnamefont
  {Jiang}}, \bibinfo {author} {\bibfnamefont {R.~I.}\ \bibnamefont
  {Gomez-Abal}}, \bibinfo {author} {\bibfnamefont {P.}~\bibnamefont {Rinke}}, \
  and\ \bibinfo {author} {\bibfnamefont {M.}~\bibnamefont {Scheffler}},\ }\href
  {\doibase 10.1103/PhysRevLett.102.126403} {\bibfield  {journal} {\bibinfo
  {journal} {Phys. Rev. Lett.}\ }\textbf {\bibinfo {volume} {102}},\ \bibinfo
  {pages} {126403} (\bibinfo {year} {2009})}\BibitemShut {NoStop}%
\end{thebibliography}


%
\end{document}